\documentclass[12pt]{article}

\usepackage{times}
\usepackage{graphicx}
\usepackage{amssymb}
\usepackage{amsthm}
\usepackage{amsmath}
\usepackage{dsfont}
\usepackage{bm}
\usepackage{mathrsfs}
\usepackage{bbold}
\usepackage{color}

\DeclareSymbolFont{rsfs}{U}{rsfs}{m}{n}
\DeclareSymbolFontAlphabet{\mathscrsfs}{rsfs}

\begin{document}

\title{Wave correlations and quantum noise in cosmology}
\author{Ulf Leonhardt\\
Department of Physics of Complex Systems,\\
Weizmann Institute of Science,\\ Rehovot 7610001, Israel
}
\date{\today}
\maketitle

\begin{abstract}
Wave noise is correlated. While it may look random in space, correlations appear in space--time, because the noise is carried by wave propagation. These correlations of wave noise give rise to fluctuation forces such as the Casimir force, they are responsible for the particle creation in the dynamical Casimir effect and in the expanding universe. This paper considers the noise correlations for light waves in non-exponentially expanding flat space. The paper determines the high-frequency asymptotics of the correlation spectrum in the conformal vacuum. These noise correlations give rise to a nontrivial vacuum energy that may appear as the cosmological constant. 
\end{abstract}

\newpage

\section{Introduction}

Explorers have mapped every corner of the Earth, but the time of exploration has only just began: $95\%$ of the current content of the universe is completely unknown. The uncharted $95\%$ are called the ``dark sector'' with $25\%$ belonging to dark matter and $70\%$ to dark energy \cite{CMBPlanck}. While there are many ideas from particle physics on the nature of dark matter, and several experimental programmes for detecting dark--matter particles \cite{LesHouches} dark energy has been an enigma \cite{LesHouchesLambda,DarkEnergy}. However, it might actually be the other way round: dark energy could be the easier problem to solve, but not as a problem of high--energy physics. Rather, it might belong to an area of low--energy physics, extrapolated to cosmological scales. In this paper I will follow up on the hypothesis \cite{Annals,London,Berechya} that dark energy, this arcane force that drives the universe apart, is a form of much more mundane forces, the van der Waals and Casimir forces, that cause ordinary things to stick. These are forces of the quantum vacuum \cite{Buhmann,Forces}.

This is not a new idea. In 1968 Zel'dovich \cite{Zeldovich} suggested that vacuum fluctuations create Einstein's cosmological constant $\Lambda$ \cite{Einstein}. Einstein's $\Lambda$ is what was later called dark energy \cite{Turner}. However, Zel'dovich's and similar suggestions \cite{WeinbergLambda} disagree with the measured value of $\Lambda$ by some 120 orders of magnitude. The idea that $\Lambda$   comes from the quantum vacuum is not new --- and seem to have failed spectacularly. What is new is a better theory of the quantum vacuum, inspired by precision measurements and manipulations of Casimir forces \cite{Lamoreaux,Levitation,CasimirEquilibrium}, by the analogy between dielectric media and space--time geometries \cite{Gordon,Quan1,Quan2,Plebanski,Schleich,Stor,GREE,LeoPhil} tried and tested in transformation optics \cite{GREE,LeoPhil,Leonhardt,Pendry} and in optical analogues of black holes \cite{Philbin,Faccio,Rubino,Genov,Bekenstein1,Bekenstein2,Drori}, and inspired by the person to whom this volume is dedicated: Michael Berry. Not only did he encourage me to pursue unconventional ideas, these ideas resonate with his work on the infinite intricacies of light \cite{Berry}.

The theory \cite{Annals,London,Berechya} is still mostly a hypothesis, but it appears to agree with astronomical data \cite{Berechya} and seems to resolve \cite{Berechya} a major inconsistency in the conventional interpretation of that data \cite{DiValentino}: the $5\sigma$ tension between the directly measured Hubble constant \cite{Riess22} and the Hubble constant inferred from the Cosmic Microwave Background \cite{CMBPlanck}. There are some $10^2$ theories to explain the Hubble tension \cite{DiValentinoTheories}. All of them require modifications of known physics --- changes to the standard model of particle physics, general relativity or the cosmological principle; all make some experimentally untested modifications, with one exception. The theory advocated here is the only one in the field rooted on experiments and relying on ``new things in old things'' --- to quote a phrase of Michael Berry.

These results are encouraging, but much more work needs to be done to prove or disprove the theory on astronomical data \cite{Berechya}, to test its physical mechanism in laboratory analogues \cite{Efrat} and also to improve the theory itself. Let me explain. The renormalized vacuum expectation value $\varepsilon_\mathrm{vac}$ of the electromagnetic energy density can be expressed such that \cite{Annals} 
\begin{equation}
\frac{4\pi G}{3c^2}\,\varepsilon_\mathrm{vac} = - \alpha_\Lambda \Delta
\label{eq:epsilon}
\end{equation}
in terms of the gravitational constant $G$, the speed of light in vacuum $c$ and the dimensionless coupling parameter $\alpha_\Lambda$. The parameter $\alpha_\Lambda$ depends on the inverse squared of the cutoff length $\ell_\Lambda$ with \cite{Annals} $\alpha_\Lambda=(9\pi)^{-1}$ if $\ell_\Lambda$ is the Planck length $\ell_\mathrm{p}=\sqrt{\hbar G/c^3}$ ($\hbar$ being the reduced Planck constant). The energy density $\varepsilon_\mathrm{vac}$ does two things: it gravitates and it generates a trace anomaly \cite{Annals,Efrat,Wald} with energy density $\varepsilon_\Lambda$ that appears as the cosmological term $\Lambda$, but is no longer constant. The total vacuum energy $\varepsilon_\Lambda+\varepsilon_\mathrm{vac}$ grows with $-4\varepsilon_\mathrm{vac}$ times the Hubble parameter \cite{Annals}. The cosmological term $\varepsilon_\Lambda$ thus accumulates $\varepsilon_\mathrm{vac}$ during the cosmic evolution, it grows with negative $\varepsilon_\mathrm{vac}$ and falls with positive $\varepsilon_\mathrm{vac}$. The cosmological constant still appears in the theory, yet not as a fundamental constant of nature but only as an integration constant \cite{Berechya} that depends on the initial conditions and presumably was zero at the beginning of time. 

The quantity $\Delta$ in the vacuum energy density (\ref{eq:epsilon}) carries the physical units of a frequency squared and depends on the nature of the quantum vacuum. In the first version \cite{Annals} of the theory $\Delta$ was found to be
\begin{equation}
\Delta = \partial_t^3 \frac{1}{H} + H \partial_t^2 \frac{1}{H} 
\label{eq:delta1}
\end{equation}
where $H$ denotes the Hubble parameter \cite{CC}. One sees from a scale analysis that $\varepsilon_\mathrm{vac}$ carries the correct order of magnitude of the cosmological constant\footnote{The argument \cite{Annals} goes as follows. According to the Friedman equation \cite{CC,LL2}  expression (\ref{eq:epsilon}) gives $\frac{1}{2}H^2$ for the realistic case of zero spatial curvature \cite{CMBPlanck}. As $H$ varies on the scale of $H$ the energy density $\varepsilon_\mathrm{vac}$ goes like $H^2$ and thus plays a role in the cosmic dynamics.}. In the second incarnation \cite{Berechya} of the theory\footnote{Actually, this was the result of my first, unpublished version of the theory.} the expression
\begin{equation}
\Delta = \partial_t^3 \frac{1}{H} 
\label{eq:delta2}
\end{equation}
was published and used to compare theory with data \cite{Berechya} assuming $\varepsilon_\mathrm{vac}$ as a perturbation of the cosmic dynamics \cite{Berechya}. While Eqs.~(\ref{eq:delta1}) and (\ref{eq:delta2}) agree on the leading term, they differ in the subdominant term. The data ruled out Eq.~(\ref{eq:delta1}) whereas Eq.~(\ref{eq:delta2}) agrees with the astronomical data with the precision of that data for exactly the Planck--scale value $\alpha_\Lambda=(9\pi)^{-1}$. However, this is only true within first--order perturbation theory; the full solution of the cosmic dynamics contains oscillatory modulations, suggesting that some vital ingredient was missing that dampens these oscillations. In this paper I hope to have identified the missing component and to have finally deduced the correct vacuum energy. The paper also clarifies the role the quantum vacuum plays in cosmology and it offers an explanation why quantum electromagnetism, and quantum electromagnetism alone, is responsible for what appears as dark energy in the current era. The heart of the problem of explaining dark energy from vacuum fluctuations is the physics of wave noise. 

\begin{figure}[h]
\begin{center}
\includegraphics[width=20pc]{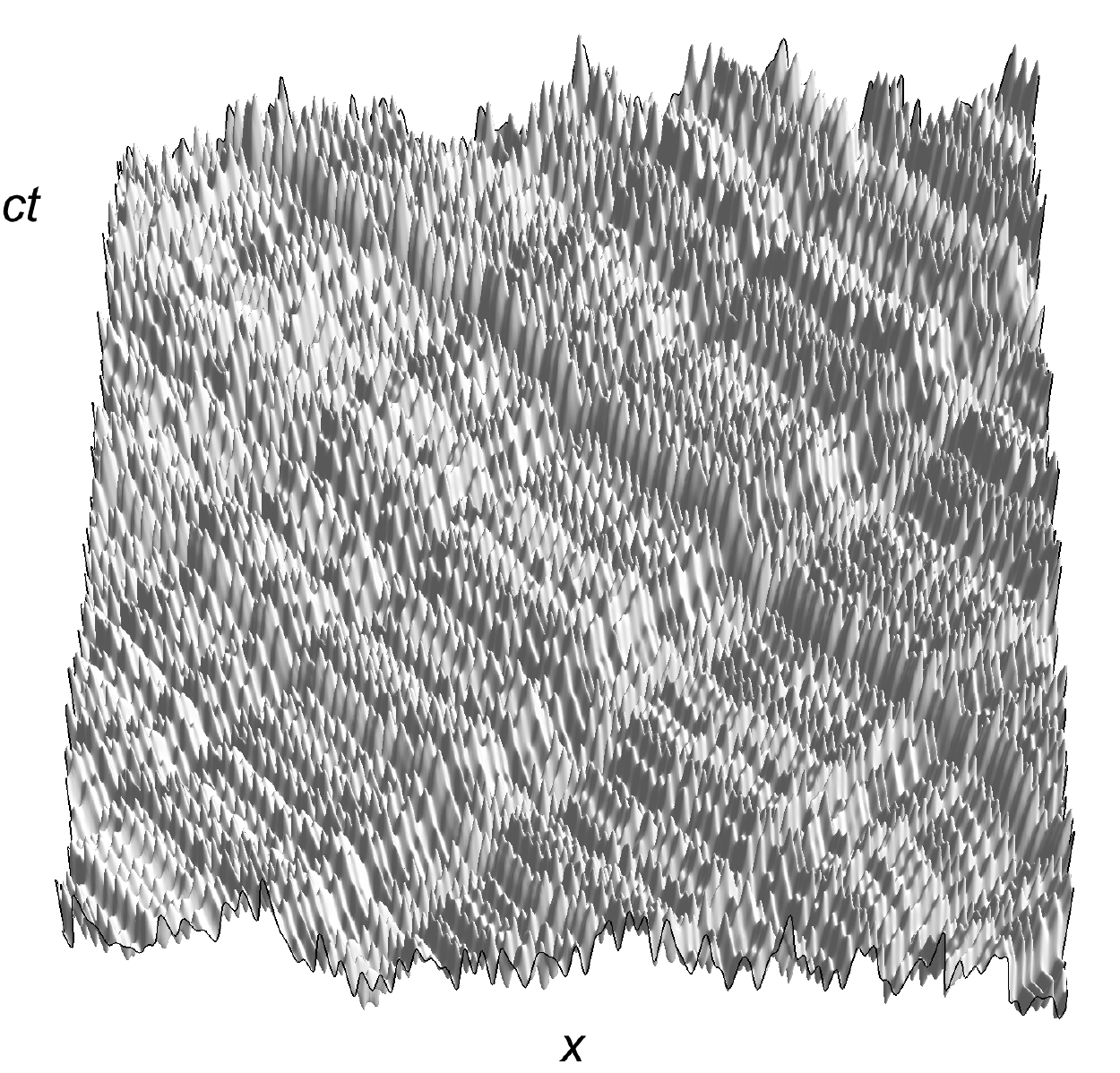}
\caption{
\small{Wave noise. Space--time diagram of waves with Gaussian noise. Although the wave field looks random in space $\{x\}$ features appear in space--time $\{ct,x\}$ following the causal cones of wave propagation (with speed $c$). For this picture 128 normalized left--moving and 128 right--moving plane waves [Eqs.~(\ref{eq:scalar0}) and (\ref{eq:norm})] with periodic boundary conditions and of random Gaussian complex coefficients were summed up. Increasing the number of waves produces finer and finer structures, but ultimately the noise field diverges, illustrating the divergence of the bare vacuum noise. 
}
\label{fig:noise}}
\end{center}
\end{figure}

Wave noise is organized. In space, it may look completely random, but in space--time patterns of correlations are clearly visible (Fig.~\ref{fig:noise}). There we see the characteristic diagonal features of wave propagation. Waves are traveling to the left or the right with the wave velocity $c/n$, and the noise they carry travels with them. If $n$ varies the noise pattern varies as well. The most dramatic of such modifications are reflections, for example at obstacles where $n$ is discontinuous. Reflected wave noise gives rise to fluctuation forces \cite{Buhmann,Forces} such as the Casimir forces \cite{Rodriguez}. If $n$ varies in time, waves may be reflected in time as well \cite{Medonca1,Medonca2}. A reflection in space is the change of sign in the wave number, in time it is a sign change in frequency. In the dynamical Casimir effect \cite{SchwingerDC,Dodonov,Wilson,Hakonen,Veccoli} these negative--frequency components correspond to newly--created particles, simply because if part of a wave of positive frequency $\omega$ is converted to $-\omega$ the energy $\hbar\omega$ of the remaining positive--frequency component must grow, particles are created. Here we focus less on the particle aspects, but rather on the amplitude correlations of wave noise. We begin with a brief review on a familiar example, the noise seen by accelerated observers \cite{Fulling,Davies,Unruh}. Then we show how this is related to the noise perceived by an observer at rest in an exponentially expanding universe \cite{GibbonsHawking} before turning to the discussion of vacuum modes in a universe of arbitrary expansion \cite{EPL}. We confirm the extension \cite{EPL} of Gibbons' and Hawking's formula for the radiation temperature \cite{GibbonsHawking} and find a new feature not present in exponential expansion: the Hawking partners appear as red--shifted thermal radiation. The multiple interference of all Hawking processes in the expanding universe gives the effective vacuum energy; to calculate it we use the Wigner function of wave noise.

\section{Uniform acceleration}

Wave noise is organized, because waves can be organized in terms of modes, and the noise appears solely in the amplitudes and phases of the mode coefficients. Consider a simple 1+1 dimensional example: a scalar wave field $\widehat{A}$ in empty Minkowski space given by the mode decomposition
\begin{equation}
\widehat{A} = \int_{-\infty}^{+\infty} \left( \widehat{a}_k A_k+ \widehat{a}_k^\dagger A_k^*\right) \mathrm{d} k
\label{eq:modes0}
\end{equation}
where the $A_k$ are the mode functions $A_k(x,t)$ describing how the modes propagate in space $x$ and time $t$. The $\widehat{a}_k$ are the mode coefficients, and only they are subject to statistical or quantum fluctuations. The mode functions should be normalized such that each mode accounts for the field of exactly one particle. This is conveniently done with the help of the scalar product \cite{LeoBook}
\begin{equation}
\left(A_1,A_2\right) = \frac{\mathrm{i}}{\hbar} \int_{-\infty}^{+\infty} \left(A_1^*\,\partial_t A_2 - A_2 \,\partial_t A_1^* \right) \mathrm{d} x 
\label{eq:scalar0}
\end{equation}
requiring 
\begin{equation}
\left(A_1,A_2\right) = \delta(k_1-k_2) \,,\quad \left(A_1^*,A_2\right) = 0 \,.
\label{eq:norm}
\end{equation}
For example, if the modes are plane waves $A_k={\cal A} \exp(\mathrm{i}kx-\mathrm{i}\omega t)$ with $\omega=c|k|$ we must require ${\cal A}^2= \hbar/(4\pi\omega)$. From the canonical commutation relations between field and momentum density then follow \cite{LeoBook} --- for Bosonic fields like the electromagnetic field --- the standard Bose commutation relations:
\begin{equation}
[\widehat{a}_{k_1},\widehat{a}_{k_2}^\dagger] = \delta(k_1-k_2) \,,\quad [\widehat{a}_{k_1},\widehat{a}_{k_2}] = 0 \,.
\label{eq:bose}
\end{equation}
The Minkowski vacuum $|0\rangle$ is the quantum state annihilated by all the plane--wave operators:
\begin{equation}
\widehat{a}_k |0\rangle = 0 \,.
\label{eq:vacuum}
\end{equation}
The Minkowski vacuum is the vacuum with respect to an observer at rest in Minkowski space. It also appears as the vacuum to observers in uniform motion, because they perceive the modes $A_k$ as plane waves as well, Doppler--shifted of course. But this is no longer true for accelerated observers \cite{Fulling,Davies,Unruh}.

\begin{figure}[t]
\begin{center}
\includegraphics[width=20pc]{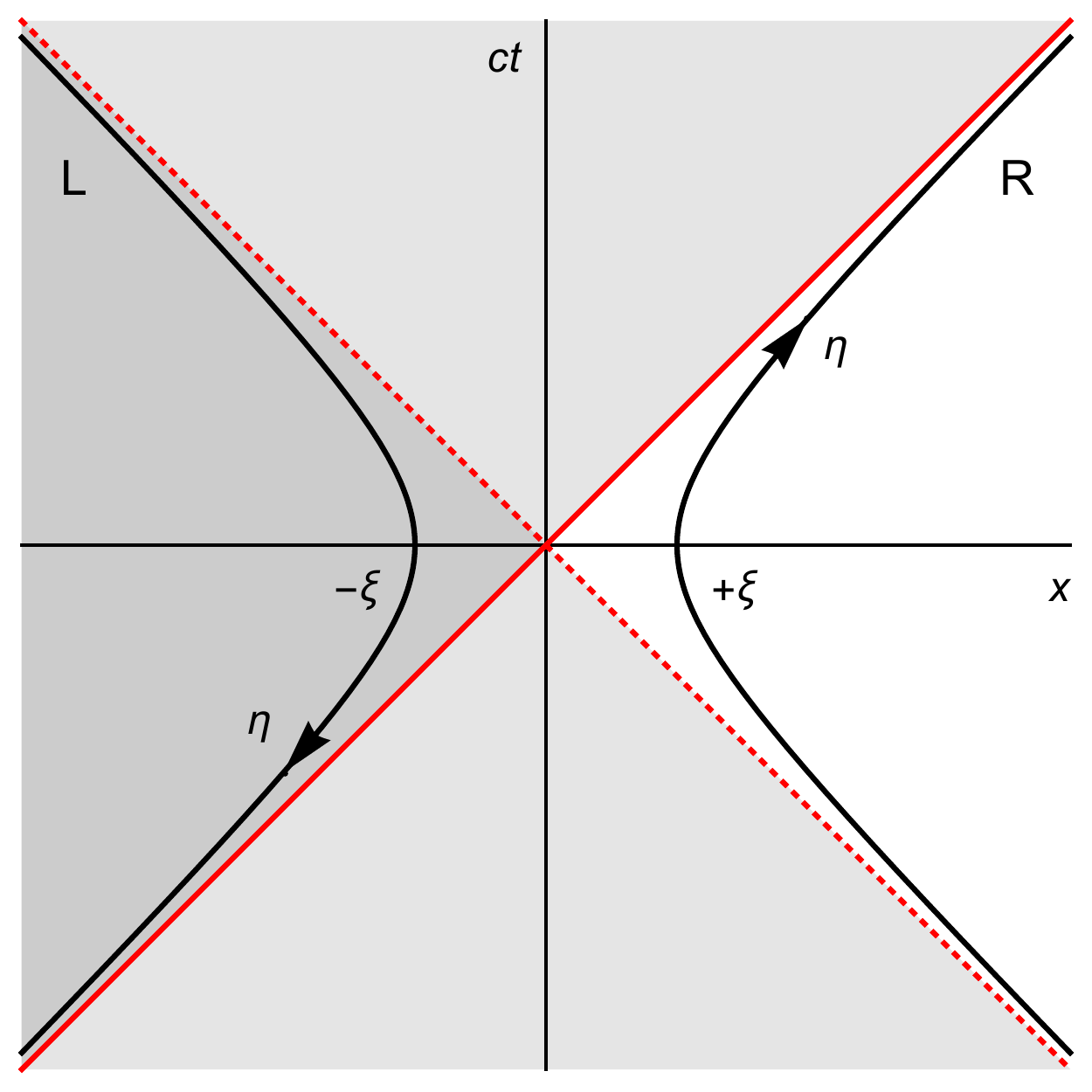}
\caption{
\small{
Accelerated observers. Space--time diagram of accelerated observers (black curves) in Minkowski space with Cartesian coordinates $x$ and $t$. The observers follow the Rindler trajectories of Eq.~(\ref{eq:rindler}) with fixed $\xi$ and variable parameter $\eta$. The acceleration is given by $c^2/\xi$ while $(\xi/c)\eta$ gives the proper time of each observer. For negative $\xi$ the parameter $\eta$ needs to run backwards (reversed arrow) as proper time always runs forwards. The observer on the right (R) is separated from the observer on the left (L) by horizons (red). Neither left-- nor right--moving light from R can reach the shaded region in L.
}
\label{fig:rindler}}
\end{center}
\end{figure}

Uniform acceleration is described by the transformation to Rindler coordinates \cite{Rindler} as follows. Suppose we write the Cartesian space--time coordinates in terms of hyperbolic polar coordinates:
\begin{equation}
x = \xi \cosh\eta\,,\quad ct=\xi\sinh\eta \,.
\label{eq:rindler}
\end{equation}
The Rindler coordinates $\{\xi,\eta\}$ cover the two wedges with $x\ge|ct|$ for $\xi\ge0$ on the right and $-x\ge|ct|$ for $\xi\le0$ on the left of the space--time diagram (Fig.~\ref{fig:rindler}). In analogy to the regular polar coordinates $\{r,\phi\}$ with spatial metric $\mathrm{d}r^2+r^2\mathrm{d}\phi^2$ we get for the hyperbolic space--time metric 
\begin{equation}
\mathrm{d}s^2 = c^2\mathrm{d}t^2 - \mathrm{d}x^2 = \xi^2\mathrm{d}\eta^2-\mathrm{d}\xi^2 \,.
\label{eq:rindlermetric}
\end{equation}
A space--time metric measures the proper time $\tau$ with increment $\mathrm{d}\tau=\mathrm{d}s/c$. In particular, as $\mathrm{d}s=\xi\mathrm{d}\eta$ for $\mathrm{d}\xi=0$, the proper time along a trajectory with fixed $\xi$ is $(\xi/c)\eta$. We can draw another conclusion from the analogy of the Rindler coordinates with polar coordinates. In space a rotation corresponds to a shift in the angle. In Minkowski space--time, a hyperbolic rotation corresponds to a Lorentz transformation to a frame moving with velocity $u$.  An infinitesimal Lorentz boost shifts the hyperbolic angle by $\mathrm{d}u/c$. A sequence of infinitesimal boosts thus draws an entire Rindler coordinate line along varying $\eta$ for $\xi=\mathrm{const}$. Now, uniform acceleration is just such a sequence of infinitesimal Lorentz transformations. We thus conclude that the Rindler line is the world line of a uniformly accelerated observer with acceleration $\mathrm{d}u/\mathrm{d}\tau = c^2/\xi$.

Consider such a uniformly accelerated observer. Suppose the observer is equipped with a spectrometer. A spectrometer consists of a spectral element to decompose the field $\widehat{A}$ into frequencies, and a detector to measure the spectral components. It is not important what the detector is. It may be a particle detector \cite{Unruh} or an amplitude detector \cite{WaterUnruh}, the physically important feature of the spectrometer is the ability to perform a frequency analysis, and there the important aspect is the fact that the spectrometer responds to its proper time $\tau$ and not to the coordinate time $t$. As $\tau = (\xi/c) \eta$ we may describe the effect of the spectrometer as a Fourier transformation with respect to $\eta$. Note, however, that for $\xi<0$ (on the left side L of the Rindler diagram of Fig.~\ref{fig:rindler}) $\eta$ needs to run backwards, since proper time always runs forwards. 

\begin{figure}[t]
\begin{center}
\includegraphics[width=20pc]{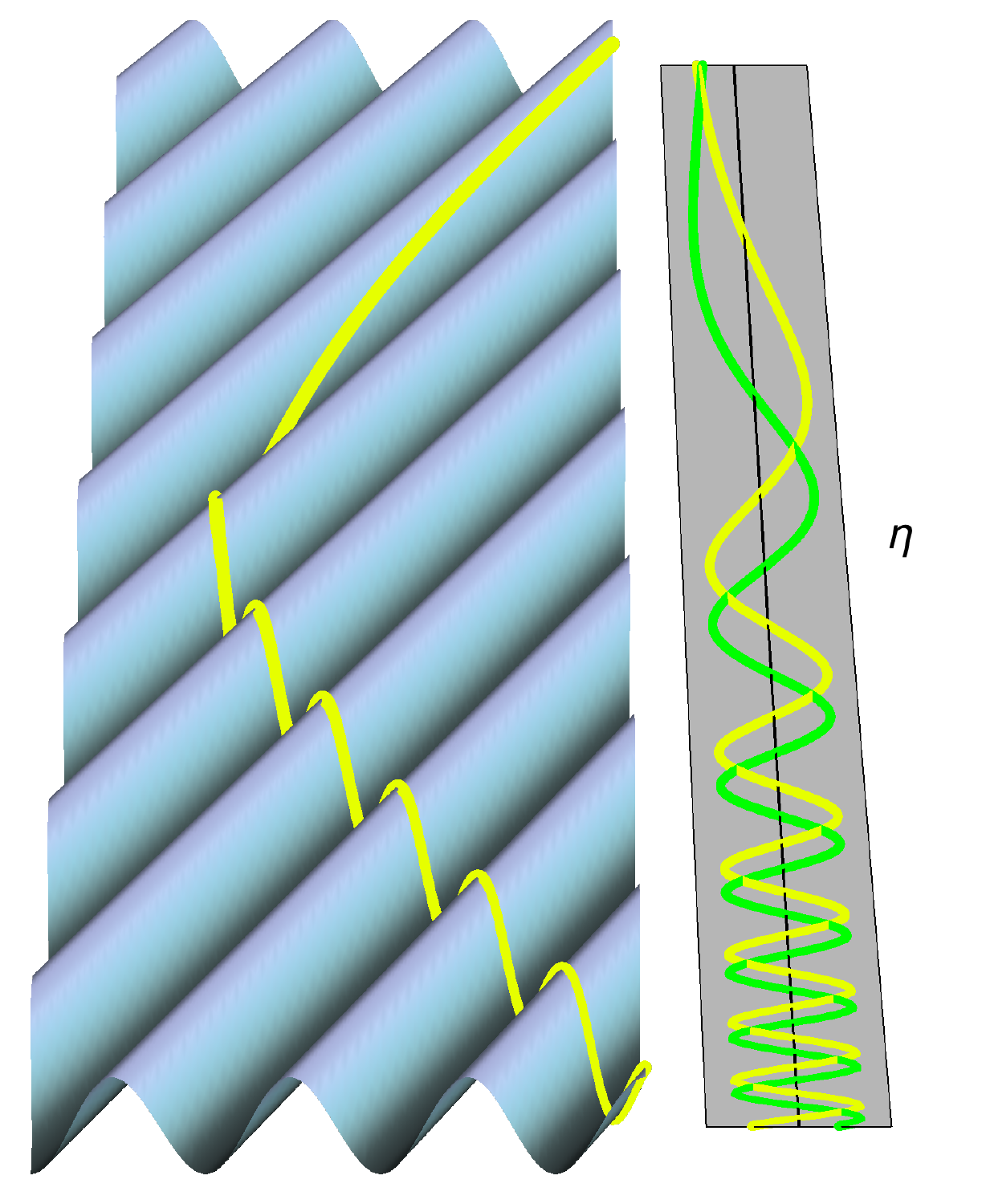}
\caption{
\small{
Plane wave. The accelerated observer (Fig.~\ref{fig:rindler}) samples noise made of plane waves with random amplitudes and phases. Each plane wave is sampled along the Rindler trajectory of Eq.~(\ref{eq:rindler}) with proper time $(\xi/c)\eta$. The panel shows the real and imaginary part of the wave sampled along the path with parameter $\eta$. Fourier analysis reveals that the positive--frequency components for $\eta$ contain negative--frequency components for $t$ enhancing the quantum noise perceived by one observer at $+\xi$ by correlations with its partner at $-\xi$ (Fig.~\ref{fig:rindler}).
}
\label{fig:sampling}}
\end{center}
\end{figure}

Imagine now a pair of accelerated observers --- one with positive $\xi$ on R and one with the exact opposite $-\xi$ on L. Figure~\ref{fig:rindler} reveals that the two observers are separated by horizons. The entire world line of observer L lies in the shadow of left-- or right--moving waves that touch observer R. But it turns out the two observers can and must communicate by sharing the same noise field. To work this out, consider the spectral components they measure:
\begin{equation}
\widetilde{A}_\mathrm{R} = \frac{1}{2\pi}\int_{-\infty}^{+\infty} \left.\widehat{A}\right|_\mathrm{R} \mathrm{e}^{\mathrm{i}\nu\eta}\,\mathrm{d}\eta \,,\quad \widetilde{A}_\mathrm{L} = \frac{1}{2\pi}\int_{-\infty}^{+\infty} \left.\widehat{A}\right|_\mathrm{L} \mathrm{e}^{-\mathrm{i}\nu\eta}\,\mathrm{d}\eta
\label{eq:fourier}
\end{equation}
in terms of the dimensionless Fourier components $\nu$. Here the R and L indicate the space--time trajectories of the two observers. They sample the plane--wave Minkowski modes (Fig.~\ref{fig:sampling}) as oscillations with phases
\begin{equation}
\varphi_\mathrm{R}=\left. k(x\mp ct) \right|_\mathrm{R} = k\xi\, \mathrm{e}^{\mp \eta} \,,\quad \varphi_\mathrm{L}=\left. k(x\mp ct) \right|_\mathrm{L} = -k\xi\, \mathrm{e}^{\mp \eta} \,.
\label{eq:phases}
\end{equation}
Now, components with positive Rindler frequencies $\nu$ may also sample negative Minkow\-ski frequencies, {\it i.e.}\ the complex--conjugated modes $A_k^*$. In fact, moving the contour of the Fourier integral by $+\mathrm{i}\pi$ on R and by $-\mathrm{i}\pi$ on L changes the sign in the phases (\ref{eq:phases}) while preserving the convergence of the Fourier integrals (\ref{eq:fourier}). We thus see that the Fourier transform of the conjugate $A_k^*$ is exactly $\mathrm{e}^{-\pi\nu}$ times the Fourier transform of $A_k$, on both sides of the Rindler wedge.

\begin{figure}[t]
\begin{center}
\includegraphics[width=20pc]{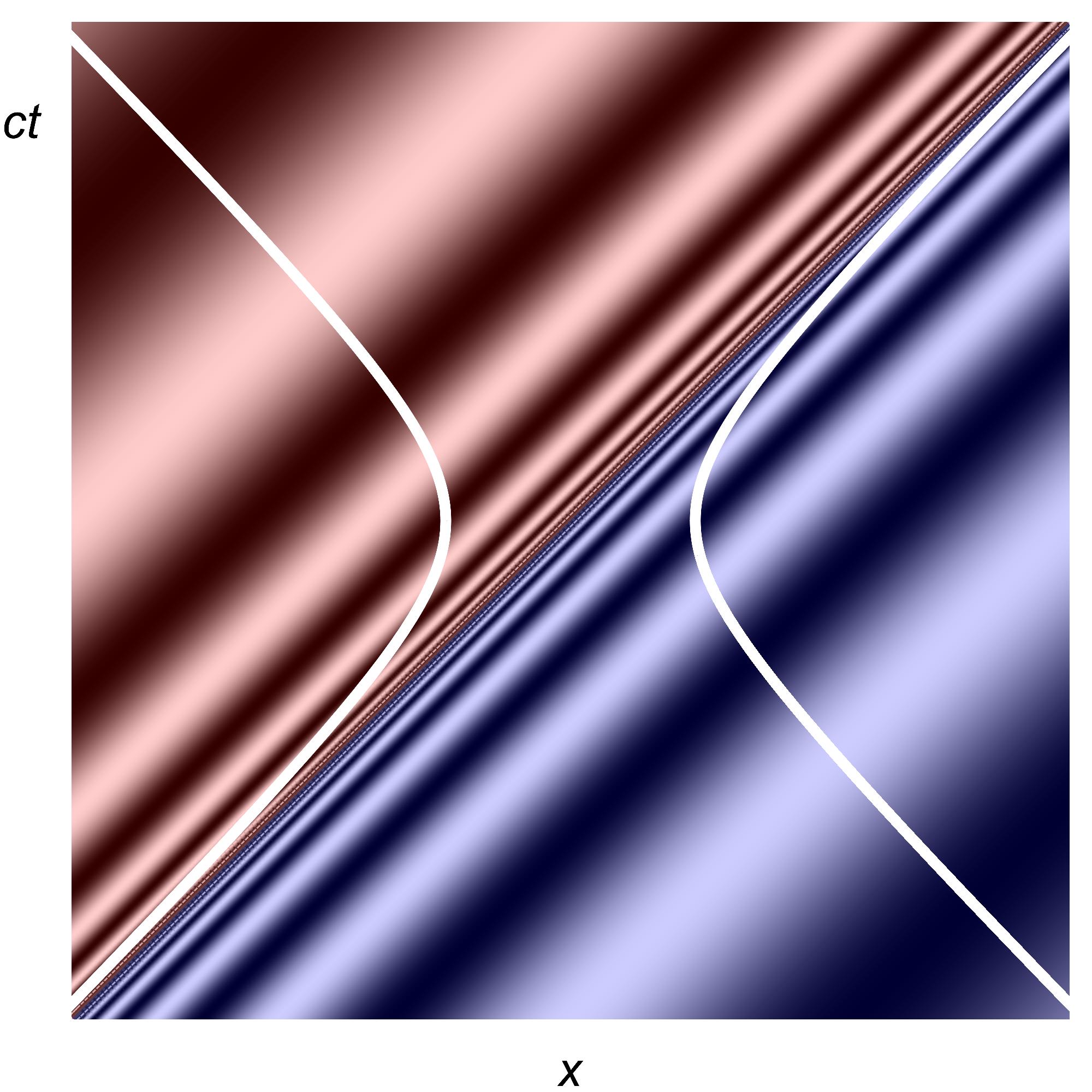}
\caption{
\small{
Rindler modes. The figure shows examples of modes that are monochromatic for the two accelerated observers (white hyperbolas, see also Fig.~\ref{fig:rindler}). For a monochromatic mode the phase increases linearly with time, but for the observers this is proper time, not coordinate time. Each accelerated observer comes in with asymptotically the speed of light and leaves asymptotically with the speed of light. For such velocities proper time ticks exponentially slowly, and so the phase grows only logarithmically. Near the horizon (Fig.~\ref{fig:rindler}) the phase diverges logarithmically [Eq.~(\ref{eq:rindlermodes})]. An exponentially small part of the wave crosses to the other side if this wave is made of a superposition of positive--norm plane waves, describing the quantum vacuum. 
}
\label{fig:rindlermodes}}
\end{center}
\end{figure}

Accelerated observers sample negative Minkowski frequencies. To see how this affects the wave noise perceived by the accelerated observers, we introduce a set of modes that are monochromatic with respect to those observers (Fig.~\ref{fig:rindlermodes}). Any mode in Minkowski space must be a superposition of left-- or right--moving waves. The left--moving waves are functions of $x_-=x+ct$ while the right--moving modes depend on $x_+=x-ct$. From $x_\pm=\xi\mathrm{e}^{\mp\eta}$ follows that the phases of monochromatic Rindler modes must be logarithmic in $x_\pm$, which means that the Rindler modes are purely imaginary powers of $x_\pm$. There we have two possibilities: $x_\pm$ or $-x_\pm$ to an imaginary power. In the first case the wave is predominately localized on the right side of the space--time diagram (Fig.~\ref{fig:rindler}), in the second case on the left side. On R we should give the Rindler wave a positive $\eta$--frequency $\nu$, {\it i.e.}\ the power $\pm\nu$ of $x_\pm$, while on L it should oscillate with $-\nu$ as $\eta$ runs backwards for forward--running proper time, which also corresponds to the power $\pm\nu$ but this time of $-x_\pm$. We thus define 
\begin{equation}
A_\nu = {\cal A} \begin{cases}
(x_\pm)^{\pm\mathrm{i}\nu} & : \nu>0 \\
(-x_\pm)^{\pm\mathrm{i}\nu} & : \nu<0 
\end{cases}
\quad\mbox{with}\quad x_\pm = x\mp ct 
\label{eq:rindlermodes}
\end{equation}
and represent the field as 
\begin{equation}
\widehat{A} = \sum_\pm \int_{-\infty}^{+\infty} \left(\widehat{a}_\nu A_\nu + \widehat{a}_\nu^\dagger A_\nu^*\right) \mathrm{d}\nu \,.
\label{eq:rindlerrepresentation}
\end{equation}
It only remains to determine the normalization factor ${\cal A}$ from Eqs. (\ref{eq:norm}). We substitute the modes (\ref{eq:rindlermodes}) into the scalar product (\ref{eq:scalar0}) with the understanding that $(A_1,A_2)$ differs from zero only when $\nu_1\sim\nu_2$. We define $\delta=\pm(\nu_2-\nu_1)$ and obtain for $\nu>0$: 
\begin{equation}
\left(A_1,A_2\right) = \frac{2c \nu}{\hbar} {\cal A}^2 \int_{-\infty}^{+\infty} (x\mp ct)^{\mathrm{i}\delta-1} \,\mathrm{d}x
= \frac{2c \nu}{\hbar} {\cal A}^2  \left(1-\mathrm{e}^{-2\pi\nu}\right) \int_0^\infty \xi^{\mathrm{i}\delta} \,\frac{\mathrm{d}\xi}{\xi} \,.
\end{equation}
Writing $\xi$ as an exponential gives $2\pi$ times the standard Fourier representation of the delta function. Defining the parameter $\zeta$ by
\begin{equation}
\tanh\zeta = \mathrm{e}^{-\pi\nu}
\label{eq:zeta}
\end{equation}
with $\cosh\zeta =(1-\mathrm{e}^{-2\pi\nu})^{-1/2}$ we thus get
\begin{equation}
{\cal A} = {\cal B} \cosh\zeta \,,\quad {\cal B}^2 = \frac{\hbar}{4\pi c\nu} \,.
\label{eq:normal}
\end{equation}
This concludes the normalization of the Rindler modes and hence the Rindler representation of the field. Only one important, subtle point remains to be discussed. 

The Rindler modes (\ref{eq:rindlermodes}) are understood to be analytic on the upper half complex plane for $x_+$ and on the lower half plane for $x_-$ such that the left side is suppressed for $\nu>0$ and the right side for $\nu<0$. In either case, the $A_\nu$ are then analytic on the lower complex plane for the time $t$. From this follows that we can always close the contour of a Fourier transformation with respect to Minkowski time $t$ for negative frequencies $\omega$ and get zero. In other words, the Rindler modes (\ref{eq:rindlermodes}) have only positive Minkowski frequencies. Therefore, they are superpositions of positive--norm Minkowski waves, and so their associated annihilation operators $\widehat{a}_\nu$ are also just superpositions of the Minkowski $\widehat{a}_k$, which implies that both share the same vacuum state $|0\rangle$.

Having established the vacuum in the Rindler representation, it is elementary to work out the spectral components seen by the two accelerated observers. We obtain from Eqs.~(\ref{eq:fourier}) and (\ref{eq:rindlerrepresentation}) for the modes (\ref{eq:rindlermodes}) with norm (\ref{eq:normal}) and $x_\pm = \xi \mathrm{e}^{\mp\eta}$ the expressions
\begin{equation}
\widetilde{A}_\mathrm{R} = {\cal B} \left(\widehat{a}_\nu\cosh\zeta + \widehat{a}_{-\nu}^\dagger\sinh\zeta\right) \,,\quad
\widetilde{A}_\mathrm{L} = {\cal B} \left(\widehat{a}_{-\nu}\cosh\zeta + \widehat{a}_\nu^\dagger\sinh\zeta\right) \,.
\label{eq:bogoliubov}
\end{equation}
We see here again that the observers sample negative--frequency components $\widehat{a}^\dagger$ with relative weight $\tanh\zeta=\mathrm{e}^{-\pi\nu}$. Representing the mode operators in terms of their real and imaginary parts (Hermitian and anti--Hermitian parts) we see that the sampled field amplitudes are connected --- the real parts are correlated and the imaginary parts anti--correlated. This means that the wave noise perceived by the observer on R is correlated with the noise perceived by observer L. Observer R is influenced by some extra randomness that comes from this connection to observer L and vice versa. That excess noise appears in the intensity as an additional contribution to the standard vacuum noise:
\begin{equation}
\langle \widetilde{A}_\mathrm{R}^\dagger  \widetilde{A}_\mathrm{R}\rangle = \langle \widetilde{A}_\mathrm{L}^\dagger  \widetilde{A}_\mathrm{L}\rangle = {\cal B}^2\left(\frac{1}{2}+\frac{1}{\mathrm{e}^{2\pi\nu}-1} \right).
\label{eq:planck}
\end{equation}
As the dimensionless $\eta$ is related to the proper time by the factor $c/\xi$, the frequencies measured in the spectrometers of the accelerated observers are related to the dimensionless $\nu$ by the same factor. We may read the $(\mathrm{e}^{2\pi\nu}-1)^{-1}$ in Eq.~(\ref{eq:planck}) as the Planck distribution $(\mathrm{e}^{\hbar\omega/k_\mathrm{B}T}-1)^{-1}$ with Unruh temperature \cite{Unruh}
\begin{equation}
k_\mathrm{B}T = \frac{\hbar c}{2\pi \xi} 
\label{eq:unruh}
\end{equation}
where $k_\mathrm{B}$ denotes Boltzmann's constant. Each one of the two observers perceives the vacuum as thermal radiation with temperature (\ref{eq:unruh}). Each one receives this extra noise, because the noise is correlated. These correlations do appear when the field amplitudes are Fourier--transformed: they are spectral correlations. In terms of particles, they appear as entangled Einstein--Podolski--Rosen pairs \cite{LeoBook}. When the spectrometer of observer R detects a particle at frequency $\omega$ so does the spectrometer of observer L (provided they are perfectly efficient). But here we are primarily concerned with amplitude noise and its cosmological implications. 

\section{Exponential expansion}

Turn now from accelerated observers in static Minkowski space to an observer at rest in the expanding universe. Consider first the conceptually simplest case: pure exponential expansion (de Sitter space \cite{deSitter}). This is the phase of the cosmic evolution we are entering at the present time and, presumably, it was the phase of inflation \cite{Inflation} just after the Big Bang (although with a much higher expansion rate then in the current era). Assume in agreement with astronomical observations \cite{Gott} that the universe is homogeneous and isotropic, and spatially flat \cite{CMBPlanck}. In this case, the space--time geometry is given by the flat--space Friedmann--Lemaitre--Robertson--Walker metric \cite{CC}:
\begin{equation}
\mathrm{d}s^2 = c^2\mathrm{d}t^2 - a^2 \mathrm{d}\bm{r}^2
\label{eq:flrw}
\end{equation}
with time--dependent scale factor $a(t)$. The scale factor describes how spatial distances expand, as the physical distance between two points at the same time $t$ is given by $a$ times the coordinate difference $r$. The spatial coordinates $\bm{r}$ are called comoving coordinates, because they do not move relative to the universe. The coordinate time $t$ is called cosmological time and, physically, it is the proper time of an observer at rest with the universe ($\mathrm{d}\bm{r}=\bm{0}$). We may introduce a new time $\tau$ called conformal time, defined as
\begin{equation}
\tau = \int \frac{\mathrm{d} t}{a} 
\label{eq:tau}
\end{equation}
such that the metric becomes conformally flat:
\begin{equation}
\mathrm{d}s^2 = a^2\left(c^2\mathrm{d}\tau^2 - \mathrm{d}\bm{r}^2\right). 
\label{eq:metric}
\end{equation}
For light rays ($\mathrm{d}s=0$) the conformal factor $a^2$ is irrelevant, and so light rays travel in conformal time and comoving space like in empty Minkowski space. As Maxwell's equations are conformally invariant \cite{LeoPhil} this remains true for full electromagnetic fields and their quantum fluctuations. We assume that the quantum vacuum is carried by plane waves in conformal time. The notation is the exact opposite as in the case of uniform acceleration: there $t$ is the time the vacuum propagates with              and $\tau$ denotes the proper time of the accelerated observer, whereas in the expanding universe the vacuum waves propagate with $\tau$ while $t$ is the proper time of the observer at rest with the universe. 

Note that the gravitational field of the universe (the space--time geometry) does distinguish a global frame --- only in this frame the metric is homogeneous and isotropic. We can of course move this frame to any point (as the universe is homogeneous) and rotate it (as it is isotropic) but the metric is different for an observer in uniform motion. Note also that although the universe is spatially flat, it is curved in space--time. One obtains for the curvature scalar \cite{LL2}
\begin{equation}
R = -\frac{6}{c^2} \left(\partial_t H + 2H^2\right) 
\label{eq:curvature}
\end{equation}
in terms of the Hubble parameter
\begin{equation}
H = \frac{\partial_t a}{a} \,.
\label{eq:hubbleparameter}
\end{equation}
In the case of exponential expansion the Hubble parameter is a constant $H_0$ such that 
\begin{equation}
a = a_0\, \mathrm{e}^{H_0 t} \,.
\label{eq:deS}
\end{equation}
In this case, the space--time curvature is negative and constant\footnote{The space--time of exponential expansion (de Sitter space) is a maximally symmetric space with constant Riemann tensor $R^{\alpha\beta}_{\mu\nu}=-(H_0/c)^2 \, (\delta^\alpha_\mu\delta^\beta_\nu-\delta^\alpha_\nu\delta^\beta_\mu)$. The negative prefactor indicates the negative curvature.} as we also see from $R=-12H_0^2/c^2$. 

\begin{figure}[h]
\begin{center}
\includegraphics[width=20pc]{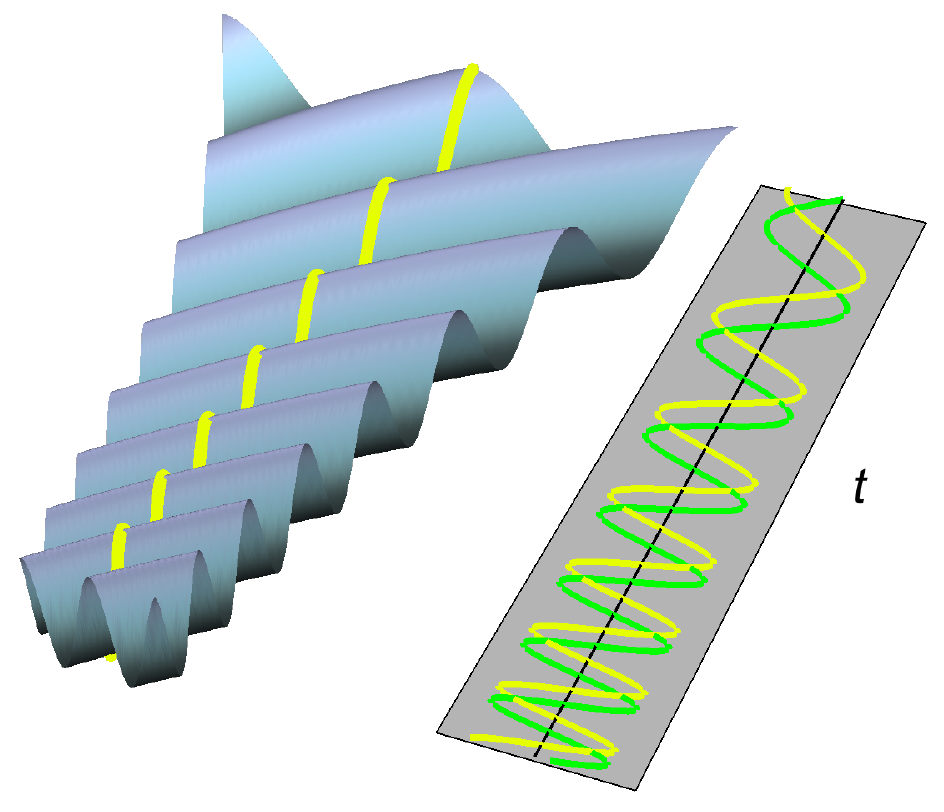}
\caption{
\small{
Exponential expansion. An observer at rest samples a plane wave in the exponentially expanding universe. The wave oscillates with conformal time [Eq.~(\ref{eq:deStau})] that differs exponentially from the proper time of the observer (the cosmological time $t$) in perfect analogy to the Minkowski wave sampled by the accelerated observer (Fig.~\ref{fig:sampling}).
} 
\label{fig:deSwaves}}
\end{center}
\end{figure}

Suppose the observer at rest with the universe samples the plane waves of the quantum vacuum (Fig.~\ref{fig:deSwaves}). They oscillate with frequencies $\Omega$ in the conformal time $\tau$ of Eq.~(\ref{eq:tau}). We obtain for the case of exponential expansion:
\begin{equation}
\tau = - \frac{1}{a H_0} \,.
\label{eq:deStau}
\end{equation}
Note that conformal time is negative and ends at $\tau=0$ in the infinite future ($t=+\infty$). The  observer samples the phase
\begin{equation}
\varphi = \Omega\tau = \frac{\Omega}{a_0 H_0}\,\mathrm{e}^{-H_0 t} \,. 
\label{eq:deSphase}
\end{equation}
This is the same phase as the one of a right--moving wave sampled by Rindler observer R (Fig.~\ref{fig:rindlermodes}). We see from Eq.~(\ref{eq:phases}) that $\Omega/(a_0H_0)$ corresponds to $k\xi$ and $H_0 t$ to the dimensionless Rindler time $\eta$. 

\begin{figure}[t]
\begin{center}
\includegraphics[width=20pc]{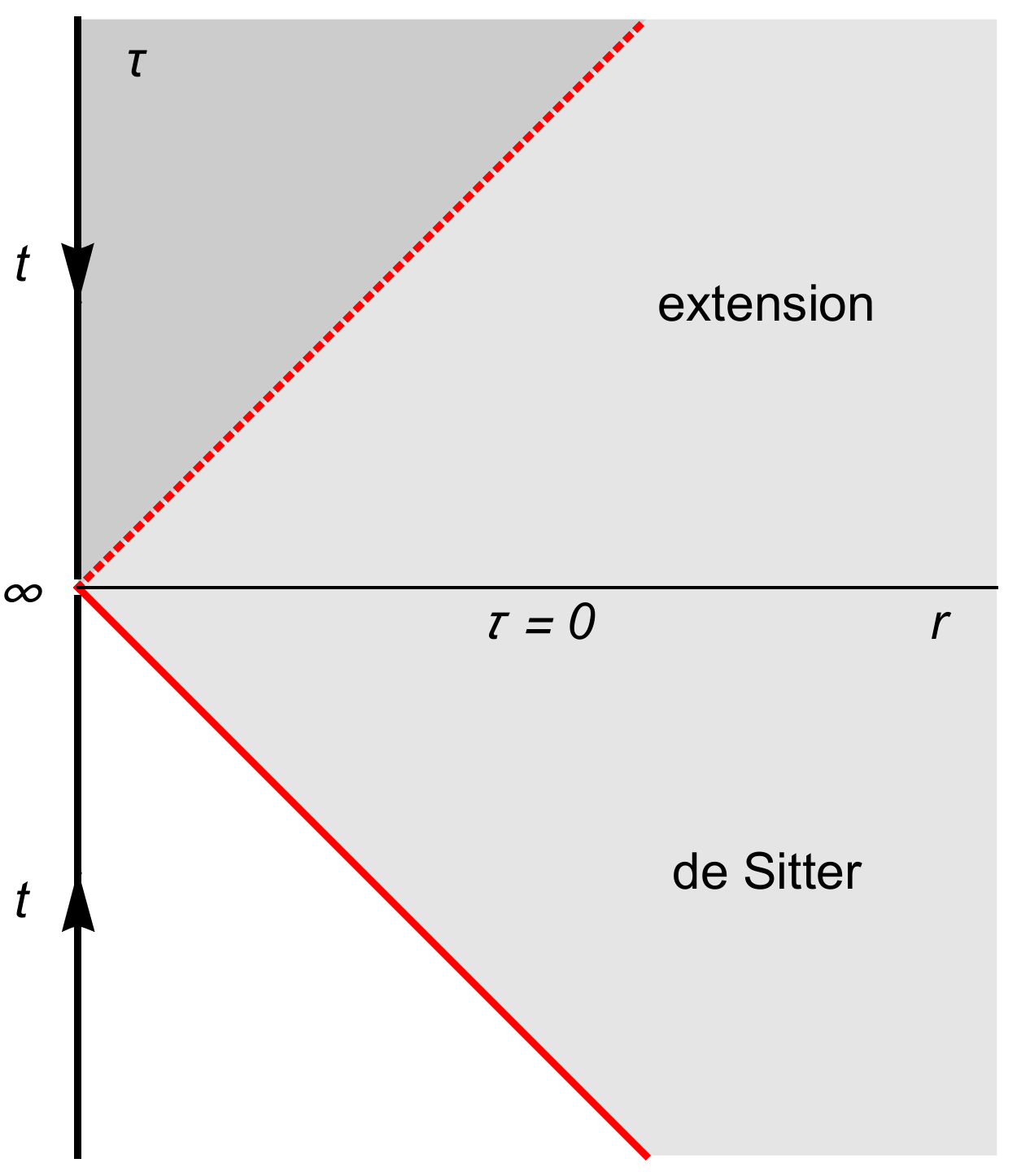}
\caption{
\small{
Extended de Sitter space. Radial space--time diagram $\{c\tau,r\}$ in conformal time $\tau$ and comoving radius $r=|\bm{r}|$. Cosmological time $t$ runs according to the arrows indicated and ends ($t=+\infty$) at the horizontal line ($\tau=0$).. Light travels along diagonal lines in the conformal diagram and may cross over to the next world, the extension, for $\tau>0$. Light beyond the horizon (red line) cannot reach the observer (black vertical line up until $t=+\infty$) before this world ends ($\tau=0$). Light coming in within the white area --- within the horizon --- leaves in the shaded area, but cannot reach the double--shaded region in the extended world, in perfect analogy to the Rindler horizon of uniform acceleration (Fig.~\ref{fig:rindler}). 
} 
\label{fig:deShorizon}}
\end{center}
\end{figure}

The observer at rest with the exponentially expanding universe thus perceives waves in the same way as the uniformly accelerated observer in Minkowski space, including the waves of the quantum vacuum. Like in uniform acceleration, the observer is surrounded by a horizon (Fig.~\ref{fig:deShorizon}). Seen in conformal time and comoving space, incoming rays outside of the radius $r_H=-c\tau$ will never arrive at the observer before the world ends in conformal time ($\tau=0$). From Eq.~(\ref{eq:deStau}) we get
\begin{equation}
r_H = \frac{c}{a H} \,.
\label{eq:horizon}
\end{equation}
Unlike the accelerated observer, there is no partner L to the observer R, at least in this universe. We may construct an artificial partner by extending de Sitter space to $\tau>0$ (similar to the Kruskal extension of the black hole \cite{Rindler}). For this we imagine another universe with infinite cosmological time related to positive conformal time by $\tau=H_0^{-1}\mathrm{e}^{-H_0 t}$. In this netherworld time runs backwards from $+\infty$ to $-\infty$ such that conformal time and light smoothly passes from one world into the other (Fig.~\ref{fig:deShorizon}). The partner observer in the netherworld is then shrouded behind a horizon (Fig.~\ref{fig:deShorizon}) from the observer in this world, in perfect analogy to uniform acceleration. In particular, we may conclude that the de Sitter observer perceives the vacuum as thermal radiation as well \cite{GibbonsHawking}. From the correspondence to the case of the accelerated observer with Unruh temperature (\ref{eq:unruh}) we obtain the Gibbons--Hawking temperature \cite{GibbonsHawking}
\begin{equation}
k_\mathrm{B}T = \frac{\hbar H_0}{2\pi} \,.
\label{eq:gh0}
\end{equation}
Exponential expansion is a clear, simple, perfectly understood case of quantum noise in cosmology, but it is largely an academic case. In reality, the universe does not expand exponentially yet nor did it in the past. Very few papers have tackled the problem beyond the case of de Sitter space \cite{EPL,Aye,Nay}, because it is a difficult problem of --- apparently --- hardly any relevance, as the Gibbons--Hawking temperature of the real universe is astronomically small ($T$ lies in the order of $10^{-29}\mathrm{K}$ for $1/H_0$ of $10\mathrm{Gy}$). But if the quantum noise of general cosmological horizons is indeed the key to understanding the cosmological constant \cite{Annals}, understand it we must. 

\section{Expanding flat space}

\begin{figure}[t]
\begin{center}
\includegraphics[width=13pc]{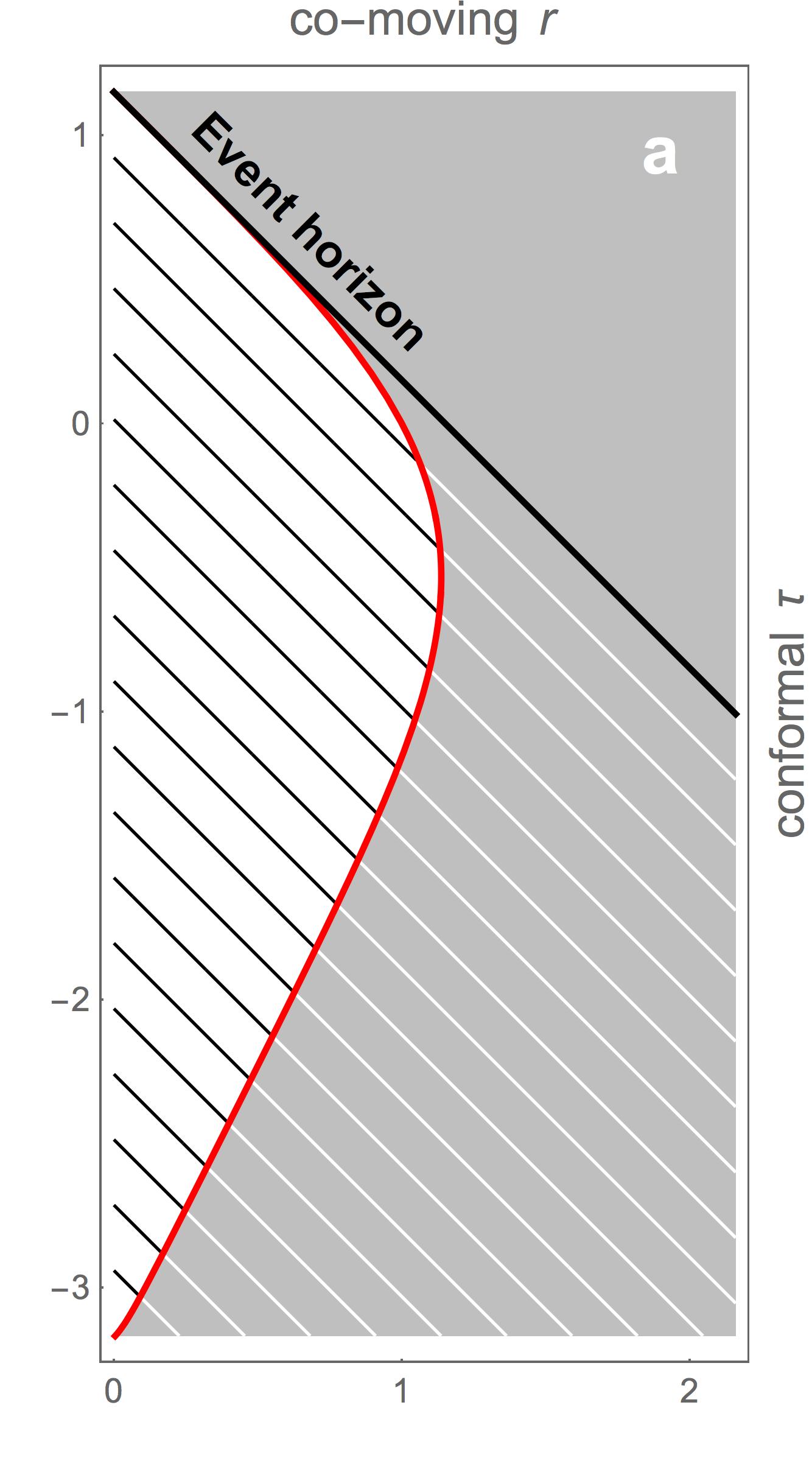}
\includegraphics[width=13.3pc]{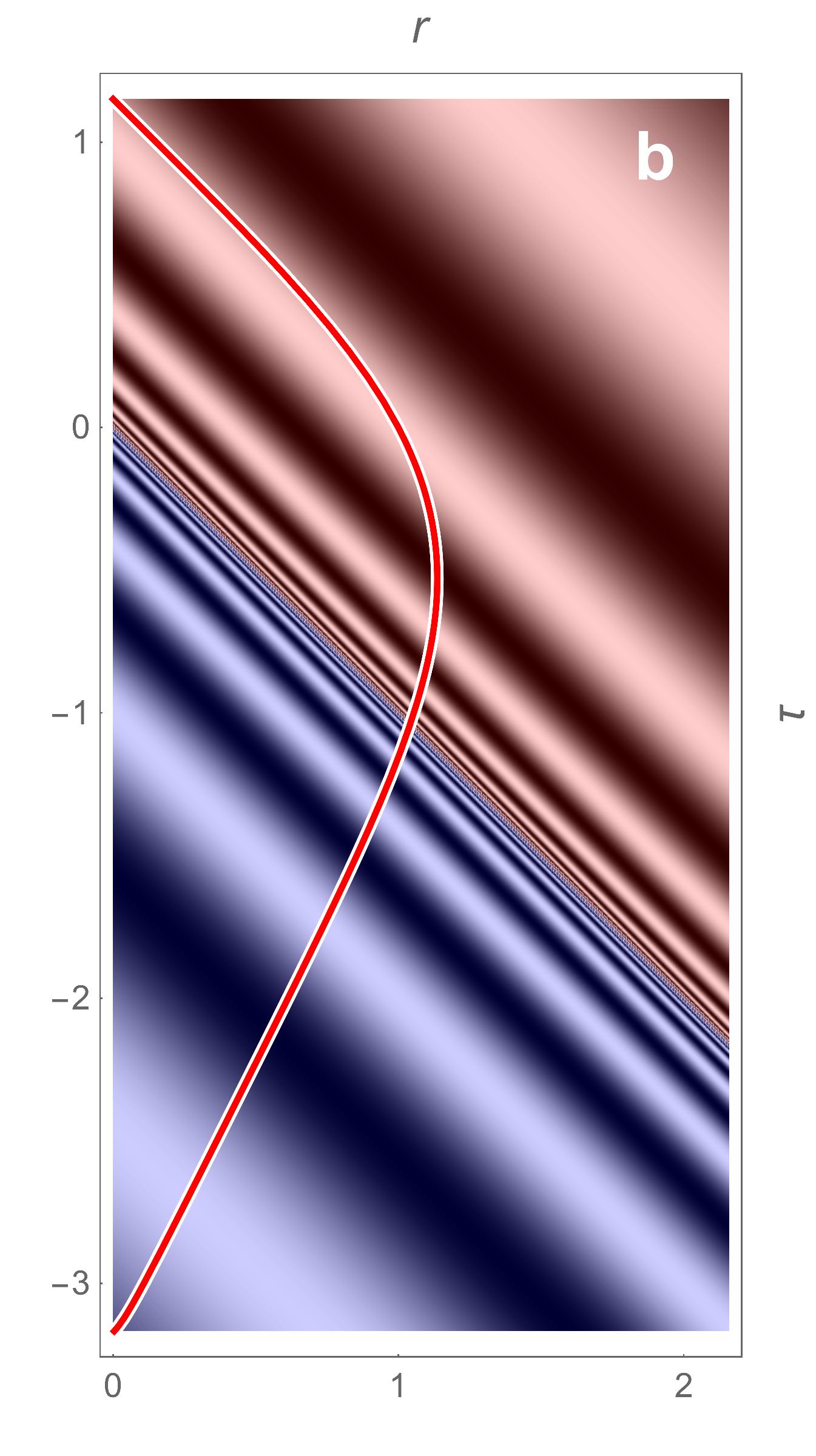}
\caption{
\small{
Cosmological horizon. Space--time diagrams of the horizon (red curve) based on actual cosmological data \cite{CMBPlanck,CC} (plotted in units $c/H_0$ with Hubble constant $H_0$). {\bf a}: in co--moving spatial coordinates $\bm{r}$ and conformal time $\tau$ light (black and white lines) propagates like in Minkowski space. The region outside the horizon is shaded in grey. Light may cross the horizon, except when, in the final stage of cosmic evolution, the horizon becomes light--like and hence a genuine event horizon. {\bf b}: vacuum modes in analogy to the Rindler modes (Fig.~\ref{fig:rindlermodes}). The modes are defined with respect to a specific time, here $\tau=0$ (the present time). The figure shows the phase pattern of the incident light only, not the outgoing light; Eq.~(\ref{eq:vacuum-modes}) describes both.
}
\label{fig:universe}}
\end{center}
\end{figure}

Apart from exponential expansion, there is no other case when an expanding flat space establishes a genuine event horizon \cite{EPL,Confusion} (Fig.~\ref{fig:universe}a). One sees this as follows. The cosmological horizon \cite{Harrison} is the spherical surface around a given point where the expansion velocity reaches the speed of light. The expansion velocity $u$ is the derivative of the proper length $\ell=ar$ with respect to cosmological time $t$. Differentiating $\ell$ gives Hubble's law, $u=H\ell$, in terms of the Hubble parameter $H$ defined in Eq.~(\ref{eq:hubbleparameter}). We see that $u$ reaches $c$ at $r_H$ of Eq.~(\ref{eq:horizon}). For the cosmological horizon to be an event horizon it needs to be light--like, parallel to light rays in the $\{c\tau,r\}$ space--time diagram, because otherwise light may cross it. Since 
\begin{equation}
\tau = \int \frac{\mathrm{d} a}{a^2H}
\label{eq:taua}
\end{equation}
the conformal time $\tau$ does only agree with $-1/(aH)$ for $H=\mathrm{const}$, {\it i.e.}\ exponential expansion, which proves that cosmological horizons are not event horizons, except in the exponential case. In fact, the light of distant galaxies and the Cosmic Microwave Background reaches us from beyond our horizon \cite{CC,Confusion}. Therefore, it is not clear from the outset how to generalize the Gibbons--Hawking formula (\ref{eq:gh0}) to the case of expanding flat space in general.\footnote{This section closely follows Ref.~\cite{EPL} but corrects an error in the conformal factor. Despite this error, the ideas and results of the paper \cite{EPL} are correct, as we show here and in Sec.~5.}

Consider light in a universe with metric (\ref{eq:flrw}). Space shall be expanding, $H>0$. For conceptual simplicity we do not start from Maxwell's equations, but rather describe each polarization component by a conformally--coupled scalar field with modes satisfying the wave equation \cite{LeoPhil,BD}:
\begin{equation}
\frac{1}{\sqrt{-g}}\,\partial_\alpha \sqrt{-g}\, g^{\alpha\beta} \partial_\beta A - \frac{R}{6}\, A  = 0
\label{eq:wave-equation}
\end{equation}
in terms of the metric tensor $g_{\alpha\beta}$, its determinant $g$ and matrix--inverse $g^{\alpha\beta}$, and the curvature scalar $R$ of Eq.~(\ref{eq:curvature}). Einstein's summation convention over repeated indices is adopted. The modes shall be normalized according to Eq.~(\ref{eq:norm}) with the scalar product \cite{BD}:
\begin{equation}
\left(A_1,A_2\right) = \frac{\mathrm{i}c}{\hbar} \int \left(A_1^*\,\partial^0 A_2 - A_2 \,\partial^0 A_1^* \right) \sqrt{-g}\,\mathrm{d}^3 x \,,\quad \partial^0 = g^{0\alpha}\partial_\alpha \,.
\label{eq:scalar}
\end{equation}
One sees from the wave equation that the scalar product (\ref{eq:scalar}) is a conserved quantity for arbitrary wave packets satisfying Eq.~(\ref{eq:wave-equation}). Writing $A$ as $A_0/a$ reduces the wave equation (\ref{eq:wave-equation}) to the free wave equation for $A_0$ with respect to the conformal time $\tau$ of Eq.~(\ref{eq:tau}), which shows that light waves propagate in the expanding universe like in free Minkowski space $\{c\tau,\bm{r}\}$ (not just light rays). We may use the plane waves 
\begin{equation}
A = ({\cal A}/a)\, \mathrm{e}^{\mathrm{i}\bm{k}\cdot\bm{r}-\mathrm{i}\omega \tau} \quad\mbox{with}\quad \omega = c|\bm{k}|\,,\quad {\cal A}^2 = \frac{\hbar}{16\pi^3\omega} 
\label{eq:planewaves}
\end{equation}
as normalized modes. We assume that the cosmological quantum vacuum is in the vacuum state (\ref{eq:vacuum}) with respect to these conformal plane waves. This cosmological vacuum is called the conformal vacuum \cite{BD}. However, as we know from the case of exponential expansion, an observer at rest may not perceive the conformal vacuum as vacuum fluctuations.  

\begin{figure}[t]
\begin{center}
\includegraphics[width=20pc]{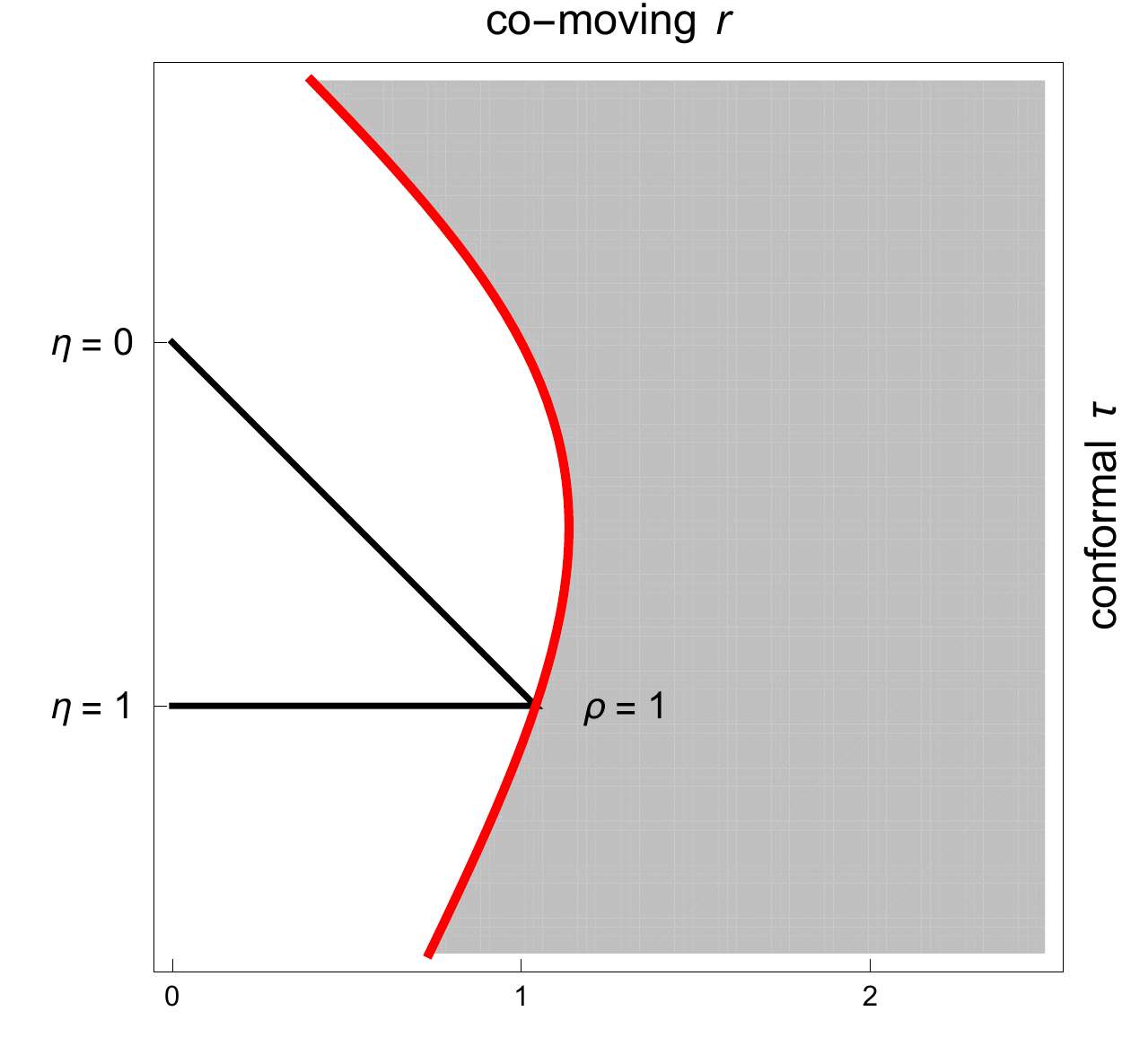}
\caption{
\small{
Characteristic events. Space--time diagram showing a part of the actual cosmological horizon (Fig.~\ref{fig:universe}a). Vacuum modes (Fig.~\ref{fig:universe}b) are established in analogy to the Rindler modes (Fig.~\ref{fig:rindlermodes}). The vacuum modes are characterized by the time parameter $\eta$ and the space parameter $\rho$ defined in Eq.~(\ref{eq:etarho}). The $\eta$ parameter runs backwards from $\eta=1$ when the vacuum mode is defined ($t=t_0$) to $\eta=0$ when the Hawking partners arrive at the origin. At the time $t_0$ ($\eta=1$) the spatial parameter reaches unity at the horizon. 
} 
\label{fig:etarho}}
\end{center}
\end{figure}

Imagine a point--like observer at rest with the expanding universe. We use spherical coordinates with the origin attached to the point of the observer. Only radial waves will matter, because all waves with higher orbital angular momentum vanish at the origin. Write the radial modes as
\begin{equation}
A = \frac{1}{\sqrt{4\pi}\,a r} A_\nu(r,\tau) \,.
\label{eq:radial}
\end{equation}
From the wave equation (\ref{eq:wave-equation}) follows that the $A_\nu$ satisfy one--dimensional wave propagation, which means that $A_\nu$ consists of a superposition of incoming and outgoing waves $f(c\tau\pm r)$. As $A$ must not diverge for $r\rightarrow 0$ we need to require $A_\nu=f(c\tau+r)-f(c\tau-r)$, the outgoing wave is the ingoing wave reflected at the focus. Inspired by the cases of uniform acceleration and exponential expansion, we wish to define modes in close analogy to the Rindler modes of Eq.~(\ref{eq:rindlermodes}). These modes can only capture the cosmological horizon at a given moment in time, {\it i.e.}\ for a given scale factor $a_0$ and corresponding Hubble parameter $H_0$. We define \cite{EPL} (Fig.~\ref{fig:universe}b) in analogy to the Rindler modes [Eq.~(\ref{eq:rindlermodes}), Fig.~\ref{fig:rindlermodes}]:
\begin{equation}
A_\nu = {\cal A} \begin{cases}
(\eta-\rho)^{\mathrm{i}\nu} - (\eta+\rho)^{\mathrm{i}\nu} & : \nu>0 \\
(\rho-\eta)^{-\mathrm{i}\nu} - (-\eta-\rho)^{-\mathrm{i}\nu} & : \nu<0 
\end{cases}
\label{eq:vacuum-modes}
\end{equation}
where $\eta$ (not to be confused with the Rindler $\eta$) and $\rho$ are defined as (Fig.~\ref{fig:etarho})
\begin{equation}
\eta = 1 + a_0 H_0 (\tau_0-\tau) \,,\quad \rho=\frac{a_0H_0}{c}\,r \,.
\label{eq:etarho}
\end{equation}
Like in the case of the Rindler modes, the modes (\ref{eq:vacuum-modes}) are analytic on the lower half $\tau$ plane. Consequently, they consist entirely of positive--frequency plane--wave modes (\ref{eq:planewaves}) and share the conformal vacuum. Let us call them vacuum modes. The phase of each of the vacuum--mode components, incoming or outgoing, is logarithmic:
\begin{equation}
\varphi = \nu \ln \left[1+a_0 H_0 (\tau_0-\tau\mp r/c)\right] \,.
\label{eq:logphase}
\end{equation}
Like the Rindler modes (Fig.~\ref{fig:rindlermodes}) the vacuum modes (\ref{eq:vacuum-modes}) are not monochromatic (Fig.~\ref{fig:universe}b); the frequency $\omega=-\partial_t\varphi$ varies in space and time. At the defining time of the modes $t_0$ we have
\begin{equation}
\left.\omega\right|_{t=t_0} = \frac{\omega_0}{1\mp u/c} \,,\quad u = H_0 \ell \,,\quad \ell = a_0 r
\label{eq:doppler}
\end{equation}
where $\omega_0$ denotes the frequency at the origin and at $t=t_0$. This frequency is related to the dimensionless parameter $\nu$ by
\begin{equation}
\omega_0 = \nu H_0 \,.
\end{equation}
Equation (\ref{eq:doppler}) shows that the vacuum modes are Doppler--shifted in the expanding universe. Incoming waves propagate against the Hubble flow $u$ and are blue--shifted, outgoing waves are red--shifted. Note that the Doppler profile (\ref{eq:doppler}) was originally used to define the modes (\ref{eq:vacuum-modes}). Here we have derived them from the analogy to the case of uniform acceleration. 

It remains to normalize the radial vacuum modes. For this we express the scalar product (\ref{eq:scalar}) in conformal time $\tau$ and spherical coordinates $\{r,\theta,\phi\}$ with metric tensor $g_{\alpha\beta}=a^2\,\mathrm{diag}(1,-1,-r^2,-r^2\sin^2\theta)$. We obtain for the radial waves (\ref{eq:radial}):
\begin{equation}
\left(A_1,A_2\right) = \frac{\mathrm{i}}{\hbar} \int_0^\infty \left(A_{\nu_1}^*\,\partial_\tau A_{\nu_2} - A_{\nu_2} \,\partial_\tau A_{\nu_1}^* \right) \mathrm{d}r \,.
\label{eq:scalar1}
\end{equation}
For the vacuum modes (\ref{eq:vacuum-modes}) with definitions (\ref{eq:etarho}) we have $\partial_\tau=-a_0H_0\,\partial_\eta$ and $a_0H_0\, \mathrm{d}r=c\,\mathrm{d}\rho$ and get
\begin{equation}
\left(A_1,A_2\right) = -\frac{\mathrm{i}c}{\hbar} \int_0^\infty \left(A_{\nu_1}^*\,\partial_\eta A_{\nu_2} - A_{\nu_2} \,\partial_\eta A_{\nu_1}^* \right) \mathrm{d}\rho \,.
\label{eq:scalar1a}
\end{equation}
We may normalize the vacuum modes at a convenient moment ($\eta=0$) as the scalar product remains constant at any time. We find exactly the same norm as for the Rindler waves, Eqs.~(\ref{eq:zeta}) and (\ref{eq:normal}). 

Finally, consider the mode overlap between the vacuum modes defined at different times. The most relevant case is the overlap between the vacuum modes at one horizon, say at $t_2$, with the modes at the previous horizon at $t_1$. By this we mean that $t_2$ is the time when the Hawking partners generated at $t_1$ arrive. The overlap tells how the modes at one instant of creating Gibbons--Hawking radiation are related to the modes at the next stage of creation. In particular, the phases between the modes are important, as the acts of creation will interfere with each other. This is because particle creation works like parametric amplification \cite{LeoBook} where the phase of the incident light determines whether particles are created or annihilated. We calculate the scalar product $(A_1,A_2)$ at time $t_2$ where $\eta_1=0$ (arrival of the partners) and $\eta_2=1$ (primary Hawking radiation). We denote the scale factors and Hubble parameters as $a_1, H_1$ and $a_2, H_2$, and use $\rho=\rho_2$ as integration variable with $\rho_1=\rho_2(a_1H_1)/(a_2 H_2)$ from Eq.~(\ref{eq:etarho}). In this way we get
\begin{equation}
\left(A_1,A_2\right) = \frac{c}{\hbar}\,(\nu_1+\nu_2)\,{\cal A}_1 {\cal A}_2 \left(\frac{a_2 H_2}{a_1 H_1}\right)^{\mathrm{i}\nu_1} \cosh^2\zeta\, I_{12}
\end{equation}
with definition (\ref{eq:zeta}) and the remaining overlap integral
\begin{eqnarray}
I_{12} &=& \int_0^\infty \rho^{-\mathrm{i}\nu_1}\, (1+\rho)^{\mathrm{i}\nu_2}\, \frac{\mathrm{d}\rho}{\rho} - \int_0^1 \rho^{-\mathrm{i}\nu_1}\, (1-\rho)^{\mathrm{i}\nu_2}\, \frac{\mathrm{d}\rho}{\rho}
\label{eq:ints}\\
& = & \frac{\Gamma(-\mathrm{i}\nu_1)}{\Gamma(-\mathrm{i}\nu_2)}\,\Gamma(\mathrm{i}\nu_2-\mathrm{i}\nu_1) -\frac{\Gamma(1+\mathrm{i}\nu_2)}{\Gamma(1-\mathrm{i}\nu_1+\mathrm{i}\nu_2)}\,\Gamma(-\mathrm{i}\nu_1)
\end{eqnarray}
in terms of Gamma functions. Note that we gave the $\nu$ an appropriate small imaginary part such that the integrals (\ref{eq:ints}) converge. The dominant contribution to the mode overlap appears for $\nu_1\rightarrow\nu_2$ where $\Gamma(\mathrm{i}\nu_2-\mathrm{i}\nu_1) \sim 1/(\mathrm{i}\nu_1-\mathrm{i}\nu_2)$. In the mode expansion $\int (\widehat{a}_\nu A_\nu+\widehat{a}_\nu^\dagger A_\nu^*)\mathrm{d}\nu$ the overlap $(A_1,\widehat{A})$ picks out a single mode with $\nu_1=\nu_2=\nu$ by Cauchy's theorem. Taking into account the normalization (\ref{eq:normal}) we arrive at the simple result: 
\begin{equation}
\widehat{a}_2 \sim \left(\frac{a_2 H_2}{a_1 H_1}\right)^{\mathrm{i}\nu} \widehat{a}_1 \,.
\label{eq:phaseshift}
\end{equation}
Therefore, to a good approximation, the coefficients of the vacuum modes at time $t_2$ are given by the mode coefficients at time $t_1$ multiplied by the characteristic logarithmic phase factor $\nu(\ln a_2 H_2 -\ln a_1 H_1)$ of the cosmological horizons. This concludes our discussion of the vacuum modes in the expanding universe.

\section{Radiating horizons}

Consider now the noise the observer perceives. The observer, at rest with the expanding universe at $r=0$, samples the field with respect to cosmological time $t$, but the field oscillates with conformal time $\tau$. In the radial vacuum modes we have organized all the superpositions of conformal plane waves the observer perceives,
such that
\begin{equation}
\left.\widehat{A}\right|_{r=0} = \int_{-\infty}^{+\infty} \left(\widehat{a}_\nu A_{0,\nu} + \widehat{a}_\nu^\dagger A_{0,\nu}^*\right) \mathrm{d}\nu 
\label{eq:expansion0}
\end{equation}
where according to Eq.~(\ref{eq:radial}) the $A_{0,\nu}$ are given by
\begin{equation}
A_{0,\nu}= \left. \frac{1}{\sqrt{4\pi}\,a r}\, A_\nu \right|_{r=0} \,.
\label{eq:a0}
\end{equation}
We obtain from expressions (\ref{eq:vacuum-modes}) and (\ref{eq:etarho}) for the modes:
\begin{equation}
\lim_{r\rightarrow 0}\frac{A_\nu}{r} = \mp 2\mathrm{i}\nu{\cal A}\,(\pm\eta)^{\pm\mathrm{i}\nu-1} \,\frac{a_0 H_0}{c}\,.
\label{eq:alimit}
\end{equation}
Consider the radiation field around two times in the cosmic evolution: near the time $t_0$ when particles are produced in the Gibbons--Hawking effect for the Hubble parameter $H_0$ and then around the time when the corresponding Hawking partners arrive, given by the condition $\eta=0$ (Fig.~\ref{fig:etarho}). The time $t_0$ is arbitrary, but for each $t_0$ a new system of modes needs to be constructed according to Eqs.~(\ref{eq:vacuum-modes}) and (\ref{eq:etarho}). Since any such system is a superposition of positive--frequency plane waves, Eq.~(\ref{eq:planewaves}), the vacuum state with respect to the mode operators $\widehat{a}_\nu$ is the cosmic vacuum, regardless of $t_0$.  

As in the cases of uniform acceleration and exponential expansion, imagine the observer as equipped with a spectrometer measuring the Fourier transformation of the field with respect to the proper time of the observer, cosmological time. Consider the Fourier transform near the time $t_0$. We write for each vacuum mode 
\begin{equation}
\widetilde{A}_{0,\nu} = \int_{-\infty}^{+\infty} A_{0,\nu}\, \mathrm{e}^{\mathrm{i}\omega t} \,\mathrm{d}t 
\end{equation}
with the understanding that the integration is performed near $t_0$. There we get from Eqs.~(\ref{eq:etarho}) and (\ref{eq:tau}):
\begin{equation}
\mathrm{d}\eta = - \frac{a_0 H_0}{a}\,\mathrm{d}t \,,\quad
t \sim - \frac{1}{H_0}\,\ln\eta \,,
\end{equation}
and hence from Eqs.~(\ref{eq:a0}) and (\ref{eq:alimit}):
\begin{equation}
\widetilde{A}_{0,\nu} = \sqrt{4\pi} \,\mathrm{i}\nu c {\cal A}\,\delta(\nu-\nu_0) \,,\quad \nu_0 = \frac{\omega}{H_0} \,.
\end{equation}
For positive frequencies $\omega$ the Fourier transform $\widetilde{A}_{0,-\nu}$ of the negative--index modes vanishes. However, like in the case of the accelerated observer, the Fourier transform of the complex conjugate negative--index modes $A^*_{0,-\nu}$ does not disappear:
\begin{equation}
\widetilde{A^*}_{0,-\nu} = \mathrm{e}^{-\pi\nu} \widetilde{A}_{0,\nu} \,.
\end{equation}
From relation (\ref{eq:zeta}) and the normalization (\ref{eq:normal}) of the vacuum modes we obtain the compact result:
\begin{equation}
\left.\int_{-\infty}^{+\infty} \widehat{A}\, \mathrm{e}^{\mathrm{i}\omega t} \,\mathrm{d}t \right|_{r=0} = \frac{\sqrt{\hbar\nu}}{c}\,\mathrm{i} \left(\widehat{a}_\nu\cosh\zeta + \widehat{a}_{-\nu}^\dagger\sinh\zeta\right) \,,\quad \nu = \frac{\omega}{H_0} \,.
\label{eq:bogoplus}
\end{equation}
The result shows that the observer, sampling the vacuum noise with respect to cosmological time around $t_0$, experiences the creation of Hawking particles \cite{EPL}, even in the case of non--exponential expansion when the cosmological horizon is not an event horizon \cite{Confusion}.

Now turn to the time $t_1$ when the Hawking partners are expected to arrive, {\it i.e.}\ when $\eta\sim0$ (Fig.~\ref{fig:etarho}). It follows from Eqs.~(\ref{eq:tau}) and (\ref{eq:etarho}):
\begin{equation}
\eta \sim -\frac{a_0 H_0}{a_1}\, t 
\end{equation}
where $a_1$ denotes the scale factor at $t_1$. Defining now $\nu_1=(a_1/a_0)(\omega/H_0)$ we thus obtain 
\begin{equation}
\widetilde{A}_{0,-\nu} = \frac{\mathrm{i}\nu}{\sqrt{\pi}}\,\frac{{\cal A}}{c} \int_{-\infty}^{+\infty} (-\eta)^{-\mathrm{i}\nu-1} \,\mathrm{e}^{-\mathrm{i}\nu_1\eta}\,\mathrm{d}\eta \sim
\mathrm{i}\sqrt{2\mathrm{i}\nu} \,\frac{{\cal A}}{c} \,(\nu_1/\nu)^{\mathrm{i}\nu}\,\mathrm{e}^{\mathrm{i}\nu}
\end{equation}
in the saddle--point approximation for $\nu\gg 1$. Similarly, for the negative--frequency Fourier--transform of the complex conjugate modes with positive index $\nu$ we get
\begin{equation}
\widetilde{A^*}_{0,+\nu} = \mathrm{e}^{-\pi\nu} \widetilde{A}_{0,-\nu} \,.
\end{equation}
Substituting these results  in the mode expansion (\ref{eq:expansion0}) we calculate the integral over the mode index in the saddle--point approximation as well. The phase of the integrand, $\varphi=\nu\ln(\nu_1/\nu)+\nu$, is stationary ($\partial_\nu\varphi=0$) for $\nu=\nu_1$. We obtain in perfect analogy to Eq.~(\ref{eq:bogoplus}):
\begin{equation}
\left.\int_{-\infty}^{+\infty} \widehat{A}\, \mathrm{e}^{\mathrm{i}\omega t} \,\mathrm{d}t \right|_{r=0} = \frac{\sqrt{\hbar\nu}}{c}\,\mathrm{i} \left(\widehat{a}_{-\nu}\cosh\zeta + \widehat{a}_\nu^\dagger\sinh\zeta\right) 
\,,\quad \nu=\frac{a_1}{a_0H_0} \,\omega \,.
\label{eq:bogominus}
\end{equation}
Like the accelerated observer [Eq.~(\ref{eq:bogoliubov})] the observer at rest with the expanding universe measures spectral correlations expressed in the Bogoliubov transformations (\ref{eq:bogoplus}) and (\ref{eq:bogominus}). These correlations appear as extra noise with Planck spectrum (\ref{eq:planck}).

\section{Cosmic cascade}

We have thus derived the thermal radiation of cosmological horizons in expanding flat space from the physical picture of wave noise (Fig.~\ref{fig:noise}). This picture reproduces the generalization \cite{EPL} of Gibbons' and Hawking's \cite{GibbonsHawking} result, Eq.~(\ref{eq:gh0}), to arbitrary expansion. In this general case $H_0$ in Eq.~(\ref{eq:gh0}) refers to the Hubble parameter (\ref{eq:hubbleparameter}) at any given time $t_0$, not required to be constant as in Gibbons' and Hawking's case \cite{GibbonsHawking} of exponential expansion, de Sitter space (Fig.~\ref{fig:deShorizon}). In addition, we also derived a new aspect of Gibbons--Hawking radiation not seen in de Sitter space. There the Hawking partners never arrive before the world ends in conformal time (Fig.~\ref{fig:deShorizon}) whereas in reality they do (Fig.~\ref{fig:etarho}). The light of distant galaxies and the Cosmic Microwave Background easily cross the cosmological horizon \cite{CC,Confusion} and so do the Hawking partners. We have found that the partners are correlated with the primary particles, Eqs.~(\ref{eq:bogoplus}) and (\ref{eq:bogominus}), for the same dimensionless frequency $\nu$. For the primary particles, $\nu$ is given by the frequency $\omega$ divided by the Hubble parameter $H_0$, which gives in the Planck spectrum (\ref{eq:planck}) the Gibbons--Hawking temperature (\ref{eq:gh0}). For the Hawking partners, $\nu$ is given by $\omega$ divided by $(a_0/a_1)H_0$ where $a_1$ denotes the scale factor at their time of arrival. This means that the Hawking partners also arrive as thermal radiation, but with the red--shifted temperature 
\begin{equation}
k_\mathrm{B}T_1 = \frac{a_0}{a_1}\,\frac{\hbar H_0}{2\pi} \,.
\end{equation}
These results are simple and intuitive, but they are still incomplete. If the present Hawking partners arrive in the future as thermal radiation, so should the Hawking partners of the past arrive in the present. Call the scale factor and Hubble parameter of the past cosmological horizon $a_{-1}$ and $H_{-1}$. The present radiation of Hawking partners should then have the temperature 
\begin{equation}
k_\mathrm{B}T_{-1} = \frac{a_{-1}}{a_0}\,\frac{\hbar H_{-1}}{2\pi} \,.
\end{equation}

\begin{figure}[t]
\begin{center}
\includegraphics[width=20pc]{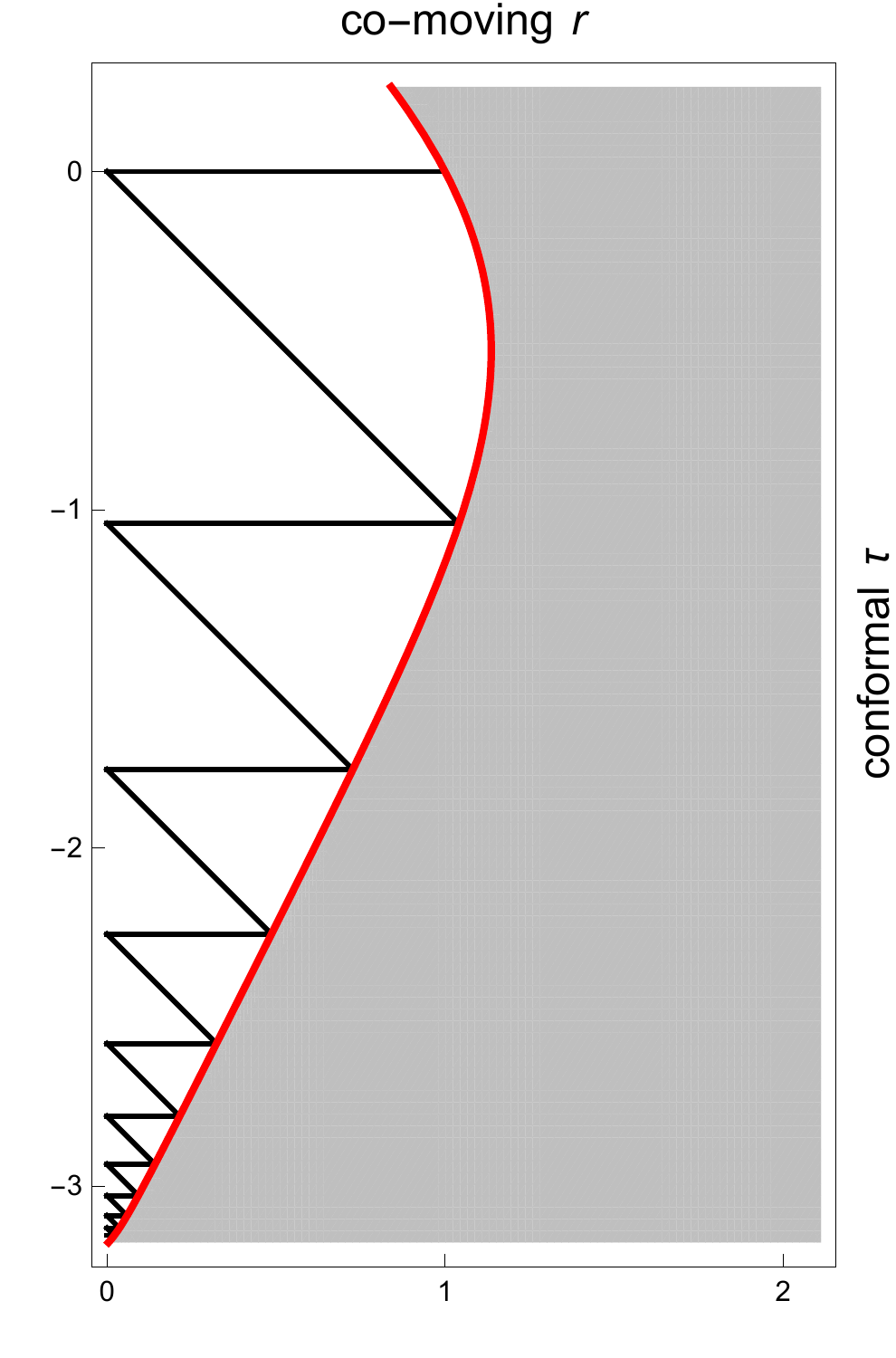}
\caption{
\small{
Cascade of horizons. In the actual universe (Fig.~\ref{fig:universe}) depicted in conformal time $\tau$ and comoving radius $r$, the Gibbons--Hawking radiation at present ($\tau=0$) depends on a cascade (zigzag line) of radiation generated by past cosmological horizons (red) curve. Depending on the relative phase, radiation is created or annihilated. The multiple interference of all creation processes gives rise to the effective Gibbons--Hawking temperature and vacuum energy density.
} 
\label{fig:cascade}}
\end{center}
\end{figure}

But neither this nor the primary temperature (\ref{eq:gh0}) is the effective temperature $T_\mathrm{eff}$ of the radiation in total, because the Hawking particles interfere with their partners. We have worked out that they have the logarithmic phase difference (\ref{eq:phaseshift}). Like in parametric amplification \cite{LeoBook} the phase of the incident radiation determines whether it gets amplified or de--amplified, whether particles are created or annihilated. The Hawking partners from the previous horizon may very well annihilate some of the Gibbons--Hawking radiation at the present, depending on the relative phase. Furthermore, the horizon before the previous horizon interferes with the particle production as well, and so does the whole cascade of past cosmological horizons (Fig.~\ref{fig:cascade}). Each horizon establishes the Bogoliubov transformation 
\begin{equation}
\widehat{b}_{\pm\nu} = \widehat{a}_{\pm\nu} \cosh\zeta + \widehat{a}_{\mp\nu}^\dagger \sinh\zeta \quad\mbox{with}\quad \tanh\zeta=\mathrm{e}^{-\pi\nu} \,.
\end{equation}
Between horizons, the modes are phase shifted according to Eq.~(\ref{eq:phaseshift}). As the frequencies relevant to the vacuum energy much exceed the Hubble parameter, we are in the regime of $\nu\gg1$ where we get for the final $\widehat{b}_\nu$ in terms of the initial vacuum mode operators $\widehat{a}_{\pm\nu}$:
\begin{equation}
\widehat{b}_\nu \sim \widehat{a}_\nu + \widehat{a}_{-\nu}^\dagger\,S\,  \mathrm{e}^{-\pi\nu}
\label{eq:bogotrans}
\end{equation}
with $S$ summing up the phase factors of the $m$--th previous horizons relative to the present one:
\begin{equation}
S = \sum_{m=1}^\infty \left(\frac{a_{-m} H_{-m}}{a_0 H_0}\right)^{2\mathrm{i}\nu} \,.
\label{eq:sum}
\end{equation}
This sum is highly oscillatory, but we are interested in the net effect of the interfering horizons, {\it i.e.}\ in the average. When averaged over $\delta\nu\sim 1$ only an exponentially small contribution will remain that, together with the primary $\mathrm{e}^{-\pi\nu}$, turns the Bogoliubov transformation (\ref{eq:bogotrans}) into 
\begin{equation}
\widehat{b}_\nu \sim \widehat{a}_\nu + \widehat{a}_{-\nu}^\dagger\, \mathrm{e}^{\mathrm{i}\Phi-\pi\omega/H_\mathrm{eff}} 
\end{equation}
with some phase $\Phi$ that does not affect vacuum correlations. The exact expression for $H_\mathrm{eff}$ we shall derive in the next section, but here we can already draw some qualitative conclusions. Since  $H_\mathrm{eff}$ depends on the history of cosmic evolution, it will introduce a memory effect in the cosmologically relevant vacuum energy. This memory of the past should remove the oscillations that would otherwise plague the cosmic dynamics. Destructive interference from past cosmological horizons may also explain why first--order perturbation theory with the primary $H$ instead of the full $H_\mathrm{eff}$ agrees so remarkably well with astronomical data \cite{Berechya}. 

It is also interesting to note that the cosmic vacuum energy vanishes within one cosmic era and thrives in the transition periods between different eras. By era we mean a period in the cosmic evolution dominated by one type of fluid with a characteristic equation of state. In the radiation--dominated era \cite{CC} the Hubble parameter $H$ goes with $a^{-2}$, in the matter--dominated era \cite{CC} $H\propto a^{-3/2}$ and during vacuum domination $H$ would become constant. Apart from the exponential expansion in the vacuum era, all other eras are characterized by a power law:
\begin{equation}
H=H_0\, a^{-\gamma}
\end{equation}
with constant $H_0$ and $\gamma>1$ (where $H_0$ denotes $H$ at $a=1$ here). The partner radiation arriving at time $t$ with scale factor $a$ and Hubble parameter $H$ originates from the past cosmological horizon the conformal time interval $\tau$ earlier, with
\begin{equation}
\tau = \int \frac{\mathrm{d}a}{a^2H} = \frac{1}{\gamma-1}\left(\frac{1}{aH}-\frac{1}{a_{-1}H_{-1}}\right) = \frac{1}{a_{-1}H_{-1}} \,,
\end{equation}
which gives
\begin{equation}
\frac{a H}{a_{-1}H_{-1}} = \frac{1}{\gamma} \,.
\end{equation}
This recurrence relation remains true for all the phases in the sum (\ref{eq:sum}) such that the sum forms a perfect harmonic Fourier series with vanishing zero--frequency component. The cycle average of such a series vanishes: the number of particles produced is exactly zero. For a power--law expansion, creation and annihilation thus cancels out exactly as the result of multiple interference between past horizons (Fig.~\ref{fig:cascade}). 

\section{Wigner function}

The interferences in the cosmic cascade of creation and annihilation at horizons (Fig.~\ref{fig:cascade}) are captured in the sum (\ref{eq:sum}). Yet this sum is difficult to evaluate and mathematically ill--defined. Let us therefore try to deduce a better formula for the effective Gibbons--Hawking temperature. The principal problem of our previous approach (Sec.~5) is the Fourier transformation. We wish to deduce the radiation spectrum as it evolves in time, and there we are interested in spectral features $\sim \exp(-2\pi/H)$ that would require an integration time in the order of $1/H$ for their accurate resolution. However, the universe also evolves on a time scale of $1/H$ and with it the Gibbons--Hawking spectrum. The two aspects, spectral accuracy and temporal resolution, appear to be mutually exclusive. Frequency and time are as mutually exclusive as position and momentum in quantum mechanics (being Fourier transforms of each other). Fortunately, there are good compromises. To give a simple example, music sheets describe tones -- frequencies --- in time; to give a sophisticated example, quantum quasiprobability distributions \cite{LeoBook,Wigner,CahillGlauber,SchleichBook} describe both position and momentum. Probably the best compromise is the Wigner function \cite{Wigner}. In quantum mechanics, the Wigner function is a partial Fourier transformation of the density matrix \cite{LeoBook}. The density matrix is a correlation function of two variables, for example two positions. The Wigner function performs a Fourier transformation with respect to the position difference as a function of the position average. In this way the Wigner function captures the momentum spectrum as a function of position. The marginal distributions (reduced probability distributions) all give the correct probability distributions of either position or momentum, or of any linear combination of the two, with perfect accuracy. This property defines the Wigner function uniquely \cite{LeoMeasuring} and explains why the Wigner function describes conjugate variables (position and momentum, time and frequency) with the highest possible precision. Here we employ the time--frequency Wigner function:
\begin{equation}
W = \frac{1}{2\pi} \int_{-\infty}^{+\infty} K(t+\theta/2,t-\theta/2)\,\mathrm{e}^{\mathrm{i}\omega \theta} \,\mathrm{d} \theta
\label{eq:wigner}
\end{equation}
of the two--time field correlation function $K$ defined as the vacuum expectation value
\begin{equation}
K = \langle \widehat{A}_1 \widehat{A}_2 + \widehat{A}_2 \widehat{A}_1 \rangle
\label{eq:Kdef}
\end{equation}
with the indices indicating the two times and positions $\{t_1, \bm{r}_1\}$ and $\{t_2, \bm{r}_2\}$. In the conformal vacuum, the electromagnetic field fluctuations propagate like in empty Minkowski space of the conformal times $\tau$ and comoving positions $\bm{r}$. We may thus use the well--known expression  of the Minkowski vacuum correlations \cite{Buhmann}:
\begin{equation}
K = \frac{1}{(2\pi)^2 s^2} \,,\quad s^2 = a_1 a_2 \left[c^2(\tau_2-\tau_1)^2-(\bm{r}_2-\bm{r}_1)^2\right]
\label{eq:K0}
\end{equation}
in terms of the Minkowski metric $s$ with reciprocal conformal factor $a_1 a_2$. Here we are interested in the spectrum measured with respect to the cosmological times $t_1$ and $t_2$ at a given point comoving with the universe:
\begin{equation}
\bm{r}_2=\bm{r}_1\,,\quad t_2=t+\theta/2\,, \quad t_1 = t-\theta/2 \,.
\end{equation}
There are several ways to derive Eq.~(\ref{eq:K0}) --- we may expand the field in terms of the plane--wave modes (\ref{eq:planewaves}) and integrate, or we may use the fact that $K$ is the real part of the analytic function $\langle \widehat{A}_1 \widehat{A}_2 \rangle$ with imaginary part given by the difference between retarded and advanced Green function, and derive  $K$ in one line from the Kramers--Kronig relation \cite{Annals}.\footnote{There is a sign error in Eq.~(50) of Ref.~\cite{Annals} and subsequent expressions, because the wrong half plane was taken in closing the integration contour. Fortunately --- thanks to another sign error --- the result (80) carries the correct sign.} Note that these vacuum correlations exist outside of the causal cone ($s<0$) as has been recently measured in quantum optics \cite{ETH}. The correlations peak at $s=0$, because electromagnetic waves propagate along light cones, including electromagnetic noise. Space--time points on the light cone ($s=0$) are thus strongly correlated. Wave noise is organized (Fig.~\ref{fig:noise}). 

Cosmology adds one subtle complication to the definition (\ref{eq:wigner}) of the Wigner function: there was a beginning of time (say at $t=0$). For a given cosmological time $t$ the Fourier time $\theta$ runs only from $-2t$ to $+2t$ in the real world. Close to the beginning, the expansion factor $a$ develops a branch point \cite{LL2} such that $a$ becomes complex in the time before, which explains \cite{LL2} why there was nothing real before the beginning of reality. The conformal time $\tau$, being defined as the integral of the inverse of $a$, inherits the branch point and ceases to be real for $|\theta|>2t$ as well. The branch points of $\tau$ are harmless in the integrand (\ref{eq:K0}), but $a$ goes to zero with some fractional power \cite{LL2}. We might be inclined to run the integral in the definition (\ref{eq:wigner}) of the Wigner function from $-2t$ to $+2t$, but the branch points of $a_1$ or $a_2$ at $\pm2t$  in the integrand (\ref{eq:K0}) would then create oscillations with period $\pi/t$ in the spectrum. The spectral oscillations average out for frequencies $\omega$ much larger than the inverse cosmic age, but like the oscillations in the cosmic cascade (Sec.~6) they obscure the subtle thermal spectrum of Gibbons--Hawking radiation. It is therefore wise to analytically continue $a$ around the beginning of time. If we lead the integral (\ref{eq:wigner}) slightly above the branch points $\pm2t$ for $\omega>0$ and slightly below for $\omega<0$ the oscillations are gone, because if we approximate $a(t)$ by some root for $t\sim 0$ we could close the integration contour on the upper half plane for $\omega>0$ and on the lower half plane for $\omega<0$ (due to the Fourier factor $\mathrm{e}^{\mathrm{i}\omega t}$) and get zero.

Having cleared the way we are now ready to calculate the Wigner function. It is wise not to use the explicit expression (\ref{eq:K0}) of the correlation function, but rather our experience with Rindler modes in expanding flat space (Sec.~4). We expand [Eq.~(\ref{eq:expansion0})] the radiation field $\widehat{A}$ in terms of the vacuum modes (\ref{eq:vacuum-modes}) defined at some arbitrary time $t_0\ge t$. The value of $t_0$ is not important, as the modes capture the conformal vacuum for all times. In the vacuum expectation value (\ref{eq:Kdef}) of $K$ the $\langle \widehat{a}^\dagger$ and $\widehat{a}\rangle$ vanish while the $\langle\widehat{a}_\nu$ and $\widehat{a}_{\nu'}^\dagger\rangle$ produce delta functions $\delta(\nu-\nu')$. For the modes at $r=|\bm{r}_2-\bm{r}_1|=0$ we apply Eqs.~(\ref{eq:a0}) and (\ref{eq:alimit}) with the normalization (\ref{eq:zeta}) and (\ref{eq:normal}), note that the negative--$\nu$ modes are reduced by the factor $\mathrm{e}^{-\pi\nu}$, and obtain the expression
\begin{equation}
K = \frac{a_0^2H_0^2}{(2\pi)^2c^2a_1\eta_1 a_2 \eta_2} \int_0^\infty 2\nu \left(\frac{1}{2}+\frac{1}{\mathrm{e}^{2\pi\nu}-1}\right) \cos\left(\nu\ln\frac{\eta_2}{\eta_1}\right)\mathrm{d}\nu \,.
\label{eq:Krind}
\end{equation}
Let us check that this formula agrees with the standard result (\ref{eq:K0}) for $K$. Formula (\ref{eq:Krind}) contains the typical Planck term $\nu(\mathrm{e}^{2\pi\nu}-1)^{-1}$ plus the contribution $\nu/2$ of the vacuum energy. We express these terms in a geometrical series:
\begin{equation}
\frac{\nu}{2} + \nu\left(\frac{1}{2}+\frac{1}{\mathrm{e}^{2\pi\nu}-1}\right) = \nu{\sum_{m=0}^\infty}' \mathrm{e}^{-2\pi m\nu}
\label{eq:vplanck}
\end{equation}
where the prime should indicate that the zeroth term is meant to be divided by 2. As
\begin{equation}
\frac{1}{4\sinh^2(z/2)} = \sum_{m=-\infty}^{+\infty} \frac{1}{(z-2\pi m\mathrm{i})^2} = -\partial_z  \sum_{m=-\infty}^{+\infty} \frac{1}{z-2\pi m\mathrm{i}} 
\label{eq:periodicpoles}
\end{equation}
we see that the term (\ref{eq:vplanck}) is the Fourier transform of $[4\sinh^2(z/2)]^{-1}$ for $\nu>0$ where we can close the integration contour on the upper half plane. Running through the pole at zero (instead of surrounding it) produces the factor $1/2$ in the vacuum term. For $\nu<0$ we close the contour on the lower half plane and get the same expression with $\nu$ replaced by $|\nu|$. From the inverse Fourier transformation then follows 
\begin{equation}
\int_0^\infty \nu \left(\frac{1}{2}+\frac{1}{\mathrm{e}^{2\pi\nu}-1}\right) \cos\nu z\, \mathrm{d}\nu = \frac{1}{4\sinh^2(z/2)} 
\end{equation}
and from this --- and definition (\ref{eq:etarho}) for $\eta$ --- we obtain Eq.~(\ref{eq:K0}). We have thus reproduced the known vacuum correlation, but only for times less than $t_0$. In the Wigner function (\ref{eq:wigner}) we must integrate from $-\infty$ to $+\infty$. Therefore we should move $t_0$ to $+\infty$. In the infinite future the expansion $a_0$ goes to infinity and $H_0$ to a finite value, and so [Eq.~(\ref{eq:etarho})] the ratio $\eta_2/\eta_1$ goes to $(\tau_\infty-\tau_2)/(\tau_\infty-\tau_1)$ while the factors $(a_0H_0)/(a\eta)$ go to $1/(\tau_\infty-\tau)$. In Eq.~(\ref{eq:Krind}) we may thus replace $\eta$ by
\begin{equation}
\eta = \tau_\infty-\tau = \int_a^\infty \frac{\mathrm{d}a}{a^2H} 
\label{eq:neweta}
\end{equation}
and remove the $a_0H_0$ altogether. We thus obtain for the thermal part of the Wigner function 
\begin{equation}
W_\mathrm{th} = \frac{1}{(2\pi)^2 c^2} \int_0^\infty \frac{\nu}{\mathrm{e}^{2\pi\nu}-1}\, \mathscrsfs{D} (\nu,\omega) \, \mathrm{d}\nu 
\label{eq:wignerthermal}
\end{equation}
with the kernel 
\begin{equation}
\mathscrsfs{D}  = \frac{1}{\pi}\,\mathrm{e}^{-\omega\sigma} \int_{-\infty}^{+\infty}\frac{1}{a_1\eta_1 a_2\eta_2} \,\cos\left(\nu\ln\frac{\eta_2}{\eta_1}\right) \mathrm{e}^{\mathrm{i}\omega \vartheta} \,\mathrm{d}\vartheta
\label{eq:d}
\end{equation}
where we have lifted the integration line in expression (\ref{eq:wigner}) by the constant positive imaginary time $\sigma$, assuming positive frequencies where, as we know, we should lead the integration contour above the branch points at the origin of physical time:
\begin{equation}
\theta = \vartheta + \mathrm{i}\sigma \quad \mbox{with}\quad\sigma>0\quad\mbox{for}\quad\omega>0 \,.
\label{eq:line}
\end{equation}
Consider de--Sitter space as a test case of our formula. In this case, $a$ grows exponentially with constant $H_0$, $\tau=-(H_0 a)^{-1}$ with $\tau_\infty=0$, and $\ln(\tau_2/\tau_1)=-H_0 (\vartheta+\mathrm{i}\sigma)$. We get 
\begin{equation}
\mathscrsfs{D} = H_0^2 \delta(\omega-H_0\nu) 
\label{eq:dSdelta}
\end{equation}
and hence a perfect Planck spectrum with Gibbons--Hawking temperature (\ref{eq:gh0}). Formula (\ref{eq:d}) thus reproduces Gibbons' and Hawking's classic result \cite{GibbonsHawking}. Consider now the realistic case of cosmic evolution, which deviates from pure exponential expansion. The kernel $\mathscrsfs{D} $ is of course independent of the integration contour (unless singularities or branch cuts are crossed) but for any given real time $t$ there will be only one imaginary time $\sigma$ when $\mathscrsfs{D}$ does approach the defining integral of a delta function in the asymptotic limit of large frequencies $\omega$ (whereas for de--Sitter space all $\sigma$ do). In the following we work out the condition when this is the case. 

But first we need to consider some realistic cosmology in order to estimate the validity of the approximation we are going to make. In the spatially flat, isotropic and homogeneous universe the square of the Hubble parameter is proportional to the energy density (by Friedman's equations \cite{CC,LL2}). For radiation (photons and neutrinos) the energy density goes with the inverse fourth power of the expansion factor $a$, because the energy falls with the inverse wavelength and hence with $a^{-1}$ and the density falls with $a^{-3}$. For matter (baryonic and dark) the energy density is essentially the rest--mass mass density multiplied by $c^2$ and $a^{-3}$. Dark energy $\Lambda$ --- being the cosmological constant --- remains constant. This gives the $\Lambda$ Cold Dark Matter ($\Lambda$CDM) model: 
\begin{equation}
H^2 = H_0^2\left(\frac{\Omega_\mathrm{R}}{a^4}+\frac{\Omega_\mathrm{M}}{a^3}+\Omega_\Lambda\right)
\label{eq:LambdaCDM}
\end{equation}
where $H_0$ denotes the Hubble constant at the present time ($a=1$) and the $\Omega_m$ describe the weights of the various contributions to the energy density with all $\Omega_m$ summing up to unity. The cosmic parameters are retrieved from the fluctuations of the Cosmic Microwave Background \cite{CMBPlanck} and are listed in Ref. \cite{CC}. For $a\gg \Omega_\mathrm{R}/\Omega_\mathrm{M}\approx 0.3\times 10^{-3}$ we can ignore the radiation contribution and enter a stage of cosmic evolution entirely dominated by matter and $\Lambda$. For describing this matter--vacuum era in the simplest possible way we change the scale of $a$ and the units of time replacing $(\Omega_\mathrm{M}/\Omega_\Lambda)^{1/3}a\rightarrow a$ and $H_0\sqrt{\Omega_\Lambda} t \rightarrow t$ such that 
\begin{equation}
H^2 = a^{-3}+1 \,.
\label{eq:LM}
\end{equation}
From $t$ being the integral of $1/(aH)$ with respect to $a$ we obtain 
\begin{equation}
a = \left(\sinh \frac{3t}{2}\right)^{2/3} \quad\mbox{and}\quad H=\coth \frac{3t}{2} \,.
\label{eq:ah}
\end{equation}
We get the conformal time
\begin{equation}
\tau = \int_0^a\frac{\mathrm{d}a}{a^2 H} = 2\sqrt{a}\,\, {}_2 F_1\left({1}/{6}, {1}/{2}, {7}/{6}, -a^3\right) \,,\quad \tau_\infty=\frac{\Gamma(1/3)\,\Gamma(7/6)}{\Gamma(3/2)} 
\label{eq:tauLM}
\end{equation}
in terms of the hypergeometric function ${}_2 F_1$ and the Gamma function $\Gamma$, and from the relationship (e.6) \cite{LL3} of the hypergeometric function:
\begin{equation}
\eta = \tau_\infty-\tau = a^{-1}\,\, {}_2 F_1\left({1}/{3}, {1}/{2}, {4}/{3}, -a^{-3}\right) \,.
\label{eq:etaLM}
\end{equation}

Consider now the curves in the complex $a$--plane where the Hubble parameter is real. For the $\Lambda$--matter model (\ref{eq:LM}) we get three curves where $H^2$ is real: straight lines going through the origin with angles $\{0,\pi/3,-\pi/3\}$. The Hubble parameter itself is real for $\infty>H^2>0$. So the curves come in from $\infty$ and end at the points where $H=\infty$ or $H=0$, which is $\{0,\mathrm{e}^{\mathrm{i}\pi/3},\mathrm{e}^{-\mathrm{i}\pi/3}\}$ for the $\Lambda$--matter stage (\ref{eq:LM}). The positive real axis corresponds to the real world with real time $t$, the $\pi/3$--line in the upper half plane corresponds to the line with positive imaginary part $\pi/3$ in the complex plane of cosmological time. In terms of the time $t+\theta/2$ in the Wigner function (\ref{eq:wigner}) it draws a line (\ref{eq:line}) parallel to the real axis with $\sigma=2\pi/3$. This is the line we are going to need in our integral (\ref{eq:d}). The $\Lambda$CDM model (\ref{eq:LambdaCDM}) has four roots of $H^2=0$ we can calculate from Ferrari's formula for the roots of quartic equations, two are real and negative, the other two complex conjugate to each other; we take the root $a_+$ on the upper half plane, for which $|a_+|=0.775$ and $\arg a_+ = \pi/3-1.09 \times 10^{-4}$. Calculating $\eta$ according to Eq.~(\ref{eq:neweta}) we find $\arg\eta_+=-\pi/3+1.34 \times 10^{-4}$. We see that $a_+\eta_+$ is real to an accuracy in the order of $10^{-5}$. 

This has consequences if we calculate the integral (\ref{eq:d}) in the saddle--point approximation for large $\omega$, because we get for the first and second derivatives of the phase $\ln(\eta_2/\eta_1)$ in the cosine:
\begin{equation}
\left.\partial_\theta \ln\frac{\eta_2}{\eta_1}\right|_{\mathrm{i}\sigma} = \left.-\mathrm{Re} \, \frac{1}{a\eta}\right|_{\mathrm{i}\sigma} \,,\quad 
\left.\partial_\theta^2 \ln\frac{\eta_2}{\eta_1}\right|_{\mathrm{i}\sigma} = \left. \frac{1}{2}\, \mathrm{Im}\left\{\frac{H}{a\eta}-\frac{1}{(a\eta)^2}\right\} \right|_{\mathrm{i}\sigma} 
\end{equation}
and so the second derivative vanishes: the integral (\ref{eq:d}) gives a delta function. In fact, for large $\omega$ only $\vartheta\sim0$ matters where we may approximate $\ln(\eta_2/\eta_1)\sim-\mathrm{i}\widetilde{\sigma}-\widetilde{H}\vartheta$ and $(a_1\eta_1 a_2\eta_2)^{-1}\sim\widetilde{H}^2$ with the definitions
\begin{equation}
\left. \widetilde{H}=\frac{1}{a\eta}\right|_{\mathrm{i}\sigma}  \,,\quad
\widetilde{\sigma} = \left.2\arg a\right|_{\mathrm{i}\sigma} \,.
\end{equation}
We thus obtain from Eq.~(\ref{eq:d}):
\begin{equation}
\mathscrsfs{D} = \mathrm{e}^{-(\omega\sigma-\nu\widetilde{\sigma})}\,\widetilde{H}^2 \delta(\omega-\widetilde{H}\nu) \,.
\label{eq:delta}
\end{equation}
For the matter--vacuum universe in our scaled units we have in particular
\begin{equation}
\widetilde{H} =  {}_2 F_1\left[{1}/{3}, {1}/{2}, {4}/{3}, \mathrm{sech}^2({3t}/{2})\right]  \,,\quad \widetilde{\sigma}=\sigma=2\pi/3 \,.
\label{eq:HLM}
\end{equation}
From Eq.~(\ref{eq:delta}) follows that, in the full $\Lambda$CDM model, the thermal part (\ref{eq:wignerthermal}) of the Wigner function (\ref{eq:wigner}) approaches the high--frequency asymptotics:
\begin{equation}
W_\mathrm{th} \sim \frac{\omega}{(2\pi)^2 c^2} \,\mathrm{e}^{-2\pi\omega/H_\mathrm{eff}} \quad\mbox{for}\quad \omega\gg\widetilde{H} 
\end{equation}
expressed in terms of the effective Hubble parameter
\begin{equation}
H_\mathrm{eff} = \frac{\widetilde{H}}{1 + \frac{1}{2\pi}(\sigma\widetilde{H} - \widetilde{\sigma})} \,.
\label{eq:Heff}
\end{equation}
The problem is solved.

\section{Summary and outlook}

We have derived the Gibbons--Hawking temperature for the standard cosmological model --- the $\Lambda$ Cold Dark Matter model --- from the physical picture of wave noise (Fig.~\ref{fig:noise}). The resulting temperature,
\begin{equation}
k_\mathrm{B}T = \frac{\hbar H_\mathrm{eff}}{2\pi} \,,
\label{eq:gh}
\end{equation}
depends on the effective Hubble parameter $H_\mathrm{eff}$ of Eq.~(\ref{eq:Heff}) with
\begin{equation}
\frac{1}{H_\mathrm{eff}} = \frac{1}{\widetilde{H}} + \frac{\sigma}{2\pi} - \frac{\widetilde{\sigma}}{2\pi\widetilde{H}} \,.
\label{eq:heff2}
\end{equation}
The effective Hubble parameter sums up the multiple interferences in the cascade of creation and annihilation at cosmological horizons (Fig.~\ref{fig:cascade}). It does it by analytic continuation of the cosmic dynamics to complex times. The parameter $\widetilde{H}$ is given by 
\begin{equation}
\left. \widetilde{H}=\frac{1}{a(\tau_\infty-\tau)}\right|_{t+\mathrm{i}\sigma/2}  
\label{eq:htilde}
\end{equation}
in terms of the scale factor $a$ and the conformal time $\tau$ evaluated at infinity and at a certain complex time $t+\mathrm{i}\sigma/2$ on the upper half plane. The real part of this complex time is the cosmological time at which Gibbons--Hawking radiation is acting at the moment, the imaginary part $\sigma/2$ needs to be determined from the requirement
\begin{equation}
\left. \mathrm{Im} \widetilde{H} \right|_{t+\mathrm{i}\sigma/2}  = 0 \,.
\end{equation}
The parameter $\widetilde{\sigma}$ is given by twice the argument of $a$ at the complex time: 
\begin{equation}
\widetilde{\sigma} = \left.2\arg a\right|_{t+{\mathrm{i}\sigma}/2} \,.
\end{equation}
Expression (\ref{eq:htilde}) generalizes the Gibbons--Hawking formula (\ref{eq:gh0}) for de--Sitter space \cite{GibbonsHawking}. In de--Sitter space \cite{deSitter} the scale factor $a$ grows exponentially as $a=\mathrm{e}^{H_0 t}$ while the conformal time $\tau$ falls as $-H_0^{-1} \mathrm{e}^{-H_0 t}$ approaching $\tau_\infty=0$ in the infinite future. The product $a(\tau_\infty-\tau)=H_0^{-1}$ clearly is constant and real for all imaginary times. In a realistic cosmological model $\sigma$ needs to be calculated. For example, in the most relevant case, the matter--vacuum dominated period of cosmic evolution, we get $\sigma=2\pi/3$ for all times $t$ (in appropriate units\footnote{Here time is measured in the inverse units of $\sqrt{\Omega_\Lambda} H_0$ where $\Omega_\Lambda H_0^2$ describes the contribution of the cosmological constant $\Lambda$ to the square of the Hubble parameter at the present time ($a=1$).}). 

\begin{figure}[h]
\begin{center}
\includegraphics[width=25pc]{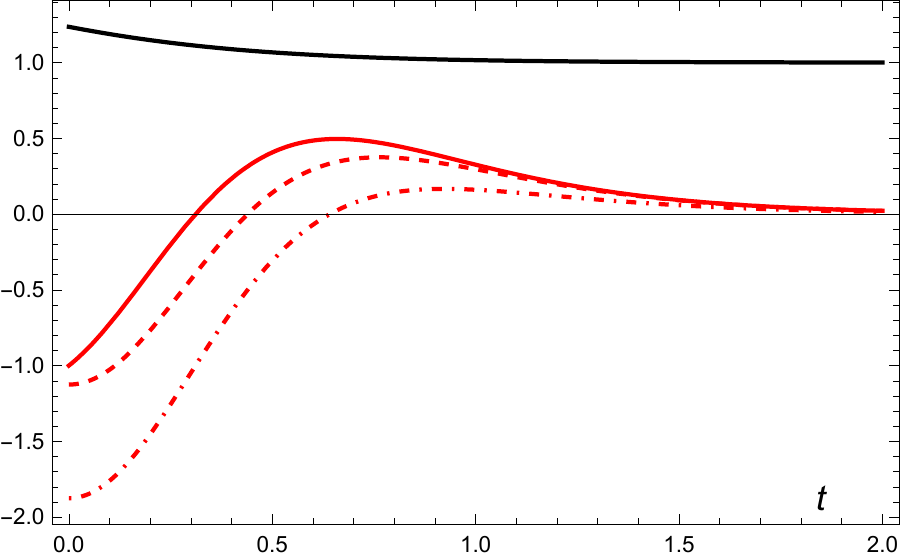}
\caption{
\small{Comparison. Black curve: $H_\mathrm{eff}$ for the matter--vacuum dominated period of cosmic evolution in scaled units. We see that $H_\mathrm{eff}$ gently falls from $1.24$ to unity for $t\rightarrow\infty$ (de Sitter space in the far future). Red curves: $\Delta$ (proportional to $-\varepsilon_\mathrm{vac}$) in scaled units. Solid curve: result of this paper, Eq.~(\ref{eq:deltaresult}). Dashed curve: $\frac{1}{6}\Delta$ obtained from Eq.~(\ref{eq:delta2}) and used, in perturbation theory, in the comparison \cite{Berechya} with astronomical data. Dashed--and--dotted curve: result of Eq.~(\ref{eq:delta1}), ruled out by the data \cite{Berechya}. The factor $\frac{1}{6}$ was chosen such that the curves have the same asymptotics for $t\rightarrow\infty$. The curves are similar, but with a different prefactor that would correspond to a different cutoff \cite{Annals}. The cutoff is a parameter of the theory, because it is not precisely known (only in its order of magnitude). It remains to be seen how the solid curve compares with  astronomical data. 
}
\label{fig:comparison}}
\end{center}
\end{figure}

The temperature (\ref{eq:gh}) lies in the order of $10^{-29}\mathrm{K}$ (at the present cosmological time) and so the particles of Gibbons--Hawking radiation are completely negligible, but the amplitude fluctuations are not --- according to Lifshitz theory \cite{Annals}. They are predicted to produce the contribution (\ref{eq:epsilon}) to the renormalized vacuum energy proportional to 
\begin{equation}
\Delta = \partial_t^3 \frac{1}{H_\mathrm{eff}} \,.
\label{eq:deltaeff}
\end{equation}
This contribution drives the cosmological term $\varepsilon_\Lambda$ \cite{Annals,London,Berechya} (but is not proportional to $\varepsilon_\Lambda$ itself). Expression (\ref{eq:deltaeff}) with effective Hubble parameter (\ref{eq:heff2}) hopefully is the final formula in a series of attempts \cite{Annals,Berechya} to determine the correct vacuum energy of expanding flat space. For the matter--vacuum dominated period we obtain in our units
\begin{equation}
\Delta = \frac{1}{\widetilde{H}^4}\left[4-8H\widetilde{H}-\left(4-\frac{26}{3}H^2\right)\widetilde{H}^2+\left(6-\frac{20}{3}H^2\right)H\widetilde{H}^3\right]
\label{eq:deltaresult}
\end{equation}
in terms of expression (\ref{eq:htilde}) at the complex time $t+\mathrm{i}\pi/3$ where $\widetilde{H}$ is real --- with $\widetilde{H}$ given by Eq.~(\ref{eq:HLM}) --- and the Hubble parameter [Eq.~(\ref{eq:ah})] that is real as well --- with $H=\tanh(3t/2)$. Figure \ref{fig:comparison} compares this result with the previous attempts for the vacuum energy, Eqs.~(\ref{eq:delta1}) and (\ref{eq:delta2}). 

Formula (\ref{eq:heff2}) depends on the history of cosmic expansion --- being determined by analytic continuation of the entire expansion. As the vacuum energy acts back on the cosmic evolution due to its gravity, it has the tendency of developing oscillations in the Hubble parameter if $\Delta$ depends on just the local values of $a$. It is hoped that the memory effect in the vacuum energy derived here will eliminate such artefacts. The multiple interference of cosmic creation and annihilation (Fig.~\ref{fig:cascade}) summed up in $H_\mathrm{eff}$ may also explain why first--order perturbation theory is remarkably good at fitting the cosmological data \cite{Berechya} while the full theory with the previous expressions would fail. 

We obtained our result (\ref{eq:heff2}) assuming that the electromagnetic vacuum noise consists of modes oscillating with conformal time (\ref{eq:tau}) whereas an observer at rest with the universe counts time as cosmological time. Furthermore we assumed, inspired by optical analogues of gravity \cite{Gordon,Quan1,Quan2,Plebanski,Schleich,Stor,GREE,LeoPhil}, that the ``medium'' of space behaves like a medium, comoving with the universe like the observer at rest. We therefore required that the spectrum of vacuum fluctuations perceived by space is the spectrum with respect to cosmological time. As the two times differ the spectrum becomes nontrivial and, as it turned out, thermal. We used the physical picture of wave noise (Fig.~\ref{fig:noise}) and the asymptology \cite{Berry} of Wigner functions \cite{SchleichBook} to work out the temperature. 

We can draw another conclusion from the picture of wave noise (Fig.~\ref{fig:noise}). Our analysis has been entirely local: we picked an arbitrary point in the spatially flat universe and considered the vacuum fluctuations at this point evolving in time. Nevertheless, the quantum vacuum is arriving from long distances away, in particular the noise of the Hawking partners. For sustaining the correlations responsible for the Gibbons--Hawking effect and hence the vacuum energy $\varepsilon_\mathrm{vac}$, perfect vacuum modes need to be formed according to Eq.~(\ref{eq:vacuum-modes}). These are superpositions of perfect, non--dispersive plane waves (\ref{eq:planewaves}) sustaining correlations across vast distances in space. Such long--range correlations cannot exist in massive fields, even for energies at the Planck scale where mass is almost irrelevant. Let us estimate the requirement for maintaining correlations. A field with particles of mass $m$ obeys the dispersion relation 
\begin{equation}
\hbar^2\omega^2 = \hbar^2 c^2 k^2 + m^2c^4 \,.
\end{equation}
Assuming $\lambda=2\pi c/\omega=\ell_\mathrm{p}$ with Planck length $\ell_\mathrm{p}$ we get for the deviation of the phase from the cosmological horizon to the point of observation: 
\begin{equation}
\delta\varphi= r_H \delta k \sim - \pi r_H \frac{\ell_\mathrm{P}} {\lambda_\mathrm{C}^2} \,,\quad \lambda_\mathrm{C} = \frac{2\pi\hbar}{mc} 
\end{equation}
where $\lambda_\mathrm{C}$ denotes the Compton wavelength. We obtain that for $r_\mathrm{H}\sim 10^{10}\mathrm{ly}$ the mass $m$ must not exceed $10^{-2}\mathrm{eV}$ for not ruining the noise correlations. There is only one field with particles of such low mass, the electromagnetic field. Gluons are massless like the electromagnetic photons, but they are short--range due to interactions with themselves. Neutrinos have masses 
$\lesssim 0.8\mathrm{eV}$ \cite{KATRIN} and are therefore probably too heavy as well. Moreover, neutrinos are fermions, and it seems questionable whether fermions can create vacuum forces. The standard vacuum fluctuations acting in the Casimir or van der Waals forces \cite{Rodriguez} are not fluctuations of particles and antiparticles, but field fluctuations. What are the physically relevant field fluctuations of fermions in the vacuum state? The need for massless bosonic fields to sustain wave correlation across cosmological distances may thus explain why only the electromagnetic field seems to contribute to the cosmological vacuum energy, as the comparison with astronomical data suggests \cite{Berechya}.

Finally, we found that for cosmological eras dominated by only one type of matter or energy the effective Gibbons--Hawking temperature is strictly zero or constant. These eras are the radiation--dominated era at the youth of the universe ($a\ll 10^{-3}$), the matter--dominated era in its middle age, and the vacuum--dominated era for the eternity to follow ($a\gg 1$). We found this by summing up the cosmic interferences, but we also see it in one glance from our analytic theory. Pure eras are described by power laws with $H=H_0\, a^{-\gamma}$, $\gamma>1$ or exponential expansion with $\gamma=0$. For a power law the conformal time $\tau$ grows with $a^{\gamma-1}$ and hence is analytic. We may close the integration contour of the Wigner function (\ref{eq:wigner}) of the vacuum correlations (\ref{eq:K0}) at infinity, get the vacuum term by integrating through the double pole at $\tau_2=\tau_1$ but zero thermal contribution. Formula (\ref{eq:htilde}) also indicates that power laws in $H$ generate zero Gibbons--Hawking temperature, because $\tau$ tends to infinity for $a\rightarrow\infty$, but the formula requires a finite $\tau_\infty$. For exponential expansion, the Gibbons--Hawking temperature (\ref{eq:gh}) is constant, and so its contribution (\ref{eq:deltaeff}) to the dynamical vacuum energy density (\ref{eq:epsilon}) vanishes as well. As the cosmological term is driven by the dynamical vacuum energy it remains constant. The vacuum energy acts only in transitions. 

This is a typical feature of the Casimir effect. In dielectrics \cite{Buhmann,Forces}, the Casimir energy thrives on differences in the dielectric properties of a medium causing forces at interfaces and boundaries. In Casimir cosmology \cite{CC}, the vacuum energy arises in the transitions between cosmic eras, changing the cosmological constant there \cite{Annals,London,Berechya}. The current era is such a transition period --- the transition from matter to vacuum domination --- and so the cosmological constant varies, which affects the Hubble constant (the Hubble parameter at the present time). The predicted variation of the Hubble constant \cite{Berechya} appears to agree with the astronomical data \cite{Riess22}, giving some empirical support to the theory presented here. Wave noise (Fig.~\ref{fig:noise}) may thus not only explain the mundane, the stickiness of the microworld, but perhaps also the arcane, the force of the macroworld that drives the universe apart. 

\section*{Acknowledgements}
Two and a half decades ago Michael Berry's work on the optical Aharonov Bohm effect inspired me to look for connections between quantum optics and general relativity, and he has been an inspiration ever since. I am most grateful to him and wish him a happy anniversary. I would also like to thank
Dror Berechya,
David Bermudez,
Nikolay Ebel,
Jonathan Kogman,
Amaury Micheli,
and
Scott Robertson
for discussions and comments on this paper. The paper has been supported by the Israel Science Foundation and the Murray B. Koffler Professorial Chair. 


\end{document}